\documentclass[twocolumn]{aastex631}

\newcommand\ysepz{\texttt{YSE PZ}}

\usepackage{threeparttable}
\usepackage{array}
\usepackage{tabularx}
\renewcommand{\thempfootnote}{\arabic{mpfootnote}}

\begin{document}

\title{Keck Infrared Transient Survey I: Survey Description and Data Release 1}

\author[0000-0002-1481-4676]{S.~Tinyanont}
\affiliation{National Astronomical Research Institute of Thailand, 260 Moo 4, Donkaew, Maerim, Chiang Mai, 50180, Thailand}
\affiliation{Department of Astronomy and Astrophysics, University of California, Santa Cruz, CA 95064, USA}

\author[0000-0002-2445-5275]{R.~J.~Foley}
\affiliation{Department of Astronomy and Astrophysics, University of California, Santa Cruz, CA 95064, USA}

\author[0000-0002-5748-4558]{K.~Taggart}
\affiliation{Department of Astronomy and Astrophysics, University of California, Santa Cruz, CA 95064, USA}

\author[0000-0002-5680-4660]{K.~W.~Davis}
\affiliation{Department of Astronomy and Astrophysics, University of California, Santa Cruz, CA 95064, USA}

\author[0000-0002-2249-0595]{N.~LeBaron}
\affiliation{Department of Astronomy, University of California, Berkeley, CA 94720-3411, USA}

\author[0000-0003-0123-0062]{J.~E.~Andrews}
\affiliation{Gemini Observatory, NSF's NOIRLab, 670 N. A'ohoku Place, Hilo, Hawai'i, 96720, USA}

\author[0000-0003-0416-9818]{M.~J.~Bustamante-Rosell}
\affiliation{Department of Astronomy and Astrophysics, University of California, Santa Cruz, CA 95064, USA}

\author[0000-0002-9830-3880]{Y.~Camacho-Neves}
\affiliation{Department of Physics and Astronomy, Rutgers, the State University of New Jersey, 136 Frelinghuysen Rd, Piscataway, NJ 08854 USA}

\author[0000-0002-7706-5668]{R.~Chornock}
\affiliation{Department of Astronomy, University of California, Berkeley, CA 94720-3411, USA}

\author[0000-0003-4263-2228]{D.~A.~Coulter}
\affiliation{Department of Astronomy and Astrophysics, University of California, Santa Cruz, CA 95064, USA}

\author[0000-0002-1296-6887]{L.~Galbany}
\affiliation{Institute of Space Sciences (ICE-CSIC), Campus UAB, Carrer de Can Magrans, s/n, E-08193 Barcelona, Spain}
\affiliation{Institut d’Estudis Espacials de Catalunya (IEEC), E-08034 Barcelona, Spain.}

\author[0000-0001-8738-6011]{S.~W.~Jha}
\affiliation{Department of Physics and Astronomy, Rutgers, the State University of New Jersey, 136 Frelinghuysen Rd, Piscataway, NJ 08854 USA}

\author[0000-0002-5740-7747]{C.~D.~Kilpatrick}
\affiliation{Center for Interdisciplinary Exploration and Research in Astrophysics (CIERA) and Department of Physics and Astronomy, Northwestern University, Evanston, IL 60208, USA}

\author[0000-0003-3108-1328]{L.~A.~Kwok}
\affiliation{Department of Physics and Astronomy, Rutgers, the State University of New Jersey, 136 Frelinghuysen Rd, Piscataway, NJ 08854 USA}

\author[0000-0003-2037-4619]{C.~Larison}
\affiliation{Department of Physics and Astronomy, Rutgers, the State University of New Jersey, 136 Frelinghuysen Rd, Piscataway, NJ 08854 USA}

\author[0000-0002-2361-7201]{J.~R.~Pierel}
\affiliation{Einstein Fellow}
\affiliation{Space Telescope Science Institute, 3700 San Martin Drive, Baltimore, MD 21218, USA}

\author[0000-0003-2445-3891]{M.~R.~Siebert}
\affiliation{Space Telescope Science Institute, 3700 San Martin Drive, Baltimore, MD 21218-2410, USA}

\author{G.~Aldering}
\affiliation{Physics Division, Lawrence Berkeley National Lab, 1 Cyclotron Rd, Berekely, CA 94720}

\author[0000-0002-4449-9152]{K.~Auchettl}
\affiliation{School of Physics, The University of Melbourne, Parkville, VIC 3010, Australia}
\affiliation{Department of Astronomy and Astrophysics, University of California, Santa Cruz, CA 95064, USA}

\author[0000-0002-7777-216X]{J.~S.~Bloom}
\affiliation{Department of Astronomy, University of California, Berkeley, CA 94720-3411, USA}
\affiliation{Lawrence Berkeley National Laboratory, 1 Cyclotron Road, MS 50B-4206, Berkeley, CA 94720, USA}

\author{S.~Dhawan}
\affiliation{Institute of Astronomy and Kavli Institute for Cosmology, University of Cambridge, Madingley Road, Cambridge CB3 0HA, UK}

\author[0000-0003-3460-0103]{A.~V.~Filippenko}
\affiliation{Department of Astronomy, University of California, Berkeley, CA 94720-3411, USA}

\author[0000-0002-4235-7337]{K.~D.~French}
\affiliation{Department of Astronomy, University of Illinois, 1002 W. Green St., Urbana, IL 61801, USA}

\author[0000-0003-4906-8447]{A.~Gagliano}
\affiliation{The NSF AI Institute for Artificial Intelligence and Fundamental Interactions}

\author[0000-0002-6741-983X]{M.~Grayling}
\affiliation{Institute of Astronomy and Kavli Institute for Cosmology, University of Cambridge, Madingley Road, Cambridge CB3 0HA, UK}

\author[0000-0002-3934-2644]{W.~V.~Jacobson-Gal\'an}
\affiliation{Department of Astronomy, University of California, Berkeley, CA 94720-3411, USA}

\author[0000-0002-6230-0151]{D.~O.~Jones}
\affiliation{Gemini Observatory, NSF's NOIRLab, 670 N. A'ohoku Place, Hilo, Hawai'i, 96720, USA}

\author[0009-0004-3242-282X]{X.~Le~Saux}
\affiliation{Department of Astronomy and Astrophysics, University of California, Santa Cruz, CA 95064, USA}

\author[0000-0002-9946-4635]{P.~Macias}
\affiliation{Department of Astronomy and Astrophysics, University of California, Santa Cruz, CA 95064, USA}

\author[0000-0001-9846-4417]{K.~S.~Mandel}
\affiliation{Institute of Astronomy and Kavli Institute for Cosmology, University of Cambridge, Madingley Road, Cambridge CB3 0HA, UK}

\author[0000-0001-5807-7893]{C.~McCully}
\affiliation{Las Cumbres Observatory 6740 Cortona Dr, Suite 102, Goleta CA 93117}

\author[0000-0003-0209-9246]{E.~Padilla~Gonzalez}
\affiliation{Department of Physics, University of California, Santa Barbara, CA 93106-9530, USA} 
\affiliation{Las Cumbres Observatory, 6740 Cortona Dr, Suite 102, Goleta, CA 93117-5575, USA}

\author[0000-0002-4410-5387]{A.~Rest}
\affiliation{Space Telescope Science Institute, 3700 San Martin Drive, Baltimore, MD 21218-2410, USA}
\affiliation{Physics and Astronomy Department, Johns Hopkins University, Baltimore, MD 21218, USA}

\author[0000-0002-7559-315X]{C.~Rojas-Bravo}
\affiliation{Department of Astronomy and Astrophysics, University of California, Santa Cruz, CA 95064, USA}

\author[0000-0001-8671-5901]{M.~F.~Skrutskie}
\affiliation{Department of Astronomy University of Virginia 530 McCormick Road, Charlottesville, VA 22904, USA}

\author[0009-0005-6323-0457]{S.~Thorp}
\affiliation{The Oskar Klein Centre, Department of Physics, Stockholm University, AlbaNova University Centre, SE 106 91 Stockholm, Sweden}
\affiliation{Institute of Astronomy and Kavli Institute for Cosmology, University of Cambridge, Madingley Road, Cambridge CB3 0HA, UK}

\author[0000-0001-5233-6989]{Q.~Wang}
\affiliation{Physics and Astronomy Department, Johns Hopkins University, Baltimore, MD 21218, USA}

\author[0000-0002-1763-2720]{S.~M.~Ward}
\affiliation{Institute of Astronomy and Kavli Institute for Cosmology, University of Cambridge, Madingley Road, Cambridge CB3 0HA, UK}

\begin{abstract}
We present the Keck Infrared Transient Survey (KITS), a NASA Key Strategic Mission Support program to obtain near-infrared (NIR) spectra of astrophysical transients of all types, and its first data release, consisting of 105 NIR spectra of 50 transients. 
Such a data set is essential as we enter a new era of IR astronomy with the {\it James Webb Space Telescope} ({\it JWST}) and the upcoming {\it Nancy Grace Roman Space Telescope (Roman)}.
NIR spectral templates will be essential to search {\it JWST} images for stellar explosions of the first stars and to plan an effective {\it Roman} SN~Ia cosmology survey, both key science objectives for mission success.
Between 2022 February and 2023 July, we systematically obtained 274 NIR spectra of 146 astronomical transients, representing a significant increase in the number of available NIR spectra in the literature. 
The first data release includes data from the 2022A semester.
We systematically observed three samples: a flux-limited sample that includes all transients $<$17~mag in a red optical band (usually ZTF $r$ or ATLAS $o$ bands); a volume-limited sample including all transients within redshift $z < 0.01$ ($D \approx 50$~Mpc); and an SN~Ia sample targeting objects at phases and light-curve parameters that had scant existing NIR data in the literature.
The flux-limited sample is 39\% complete (60\% excluding SNe~Ia), while the volume-limited sample is 54\% complete and is 79\% complete to $z = 0.005$.  
All completeness numbers will rise with the inclusion of data from other telescopes in future data releases.
Transient classes observed include common Type Ia and core-collapse supernovae, tidal disruption events (TDEs), luminous red novae, and the newly categorized hydrogen-free/helium-poor interacting Type Icn supernovae.
We describe our observing procedures and data reduction using \texttt{Pypeit}, which requires minimal human interaction to ensure reproducibility. 

\end{abstract}

\newpage


\section{Introduction} \label{sec:intro}
Studies of astrophysical transients provide insights into most subfields of astronomy.
Exploding white dwarfs in binary systems containing low-mass stars result in Type Ia supernovae (SNe).
These thermonuclear explosions produce most of the cosmic iron-group elements (IGEs).
Even with their uncertain explosion mechanism and nature of their progenitor system, SNe~Ia serve as standardizable candles used to discover the accelerating expansion of the Universe \citep{phillips1993, Riess98:lambda,Perlmutter99}.
Refining cosmological measurements using SNe~Ia is a primary mission of the \textit{Nancy Grace Roman Space Telescope} \citep[\textit{Roman;}][]{spergel2015}. 
At the high-mass end, diverse explosions of massive stars in core-collapse (CC) SNe paint a complicated picture of the evolution of the most consequential stellar constituents of the Universe. 
Massive stars evolve quickly, providing the first chemical enrichment to the nascent Universe.
These SNe, resulting from Population~III stars, could be the most distant objects discovered by the \textit{James Webb Space Telescope} (\textit{JWST}) and could be critical to the epoch of reionization \citep{Pan12:reionization, Whalen13:disc}.
On a galactic scale, tidal disruption events (TDEs; e.g., \citealp{rees88, evans89}) allow us to probe the innermost regions of distant galaxies near their otherwise quiescent supermassive black hole as it swallows an encroaching star.  

Our knowledge of transient events has greatly expanded in the past two decades, thanks to the advent of wide-field untargeted transient surveys that have continuously revealed a population of transients at all luminosity scales unrelated to bright massive galaxies in the local Universe. 
Current ongoing surveys of this nature include the All-Sky Automated Survey for Supernovae (ASAS-SN; \citealp{shappee2014}), the Asteroid Terrestrial-impact Last Alert System (ATLAS; \citealp{tonry2018, smith2020}), the Young Supernova Experiment (YSE; \citealp{jones2021}), and the Zwicky Transient Facility (ZTF; \citealp{bellm2019, masci2019}).

These visible-light surveys are enabled by the advancement in semiconductor technology, which produces silicon-based visible-light detectors that are larger, cheaper, and more efficient.
The accessibility of visible-light detectors also drives the ubiquity of spectroscopic follow-up facilities in the visible band on telescopes large and small around the globe. 
These spectroscopic datasets are the key to deciphering the nature of different types of transients. 
As of July 2023, there are almost 50,000 optical spectra of transients publicly available on WISeREP \citep{yaron2012}\footnote{\url{https://www.wiserep.org/}}, with hundreds more being obtained every month.  
This vast repository of optical spectra provides a library against which new observations and theoretical models can be compared. %

Most classes of transients emit primarily in visible light, especially when they are near maximum brightness. 
Many of the strongest atomic transition lines that allow us to probe the dynamics and chemistry of an explosion are also in the optical. 
However, focusing on the optical part of the electromagnetic spectrum alone, we would miss many crucial features of transients. 

Infrared (IR) light contains unique information from astrophysical transients. 
As they expand and cool, their spectral energy distribution (SED) shifts from peaking in the optical into the IR.
As such, IR observations are \emph{crucial} to track late-time bolometric light curves that can reveal the nature of the power source of the explosion or delayed interaction with a distant CSM. 
The IR is rich with spectral features from molecules and dust grains that form in many types of transients with suitable physical and chemical conditions, especially in CCSNe \citep[e.g.,][]{spyromilio1988, gerardy2000, gall2011, sarangi2018, rho2018, tinyanont2019,rho2021, shahbandeh2023, tinyanont2023}, but also in Type Iax SN\,2014dt \citep{fox2016} and recently in peculiar super-Chandrasekhar Type Ia SNe \citep{siebert2023, kwok2023}.
Dust formation in massive stars is likely responsible for the dust content in the early universe \citep[e.g.,][]{gall2018, shahbandeh2023}. 
Helium's strongest (1.0830~$\mu$m) and least contaminated (2.0581~$\mu$m) lines are in the IR, and observations of them allow us to unambiguously measure or constrain the helium mass in stripped-envelope (SE) SNe \citep[e.g.,][]{dessart2020}. 
There are also several lines from iron-group and intermediate-mass elements that are crucial for probing the products of explosive nucleosynthesis \citep[e.g.,][]{jerkstrand2016, jerkstrand2017, mazzali2019}. 
Furthermore, IR light is much less absorbed than optical by interstellar dust, allowing for a more homogeneous study of transients that can reside in denser environments in their host galaxies that are heavily obscured at shorter wavelengths \citep[e.g.,][]{kasliwal2017, jencson2019}.

The NIR is also promised to be revolutionary for SN Ia cosmology.
Cosmologists use a spectral model that describes the temporally evolving SED of an SN~Ia to interpret light-curve data, avoiding $K$-corrections.  
The SED is adjusted depending on SN parameters such as decline rate and color, and through this transfer function \citep[e.g.,][]{tripp1998}, distances are determined.  
Errors in the spectral model propagate to distance errors and biases, some of which will depend on redshift as filters shift through the rest frame, resulting in what is currently the largest systematic uncertainty for SN~Ia cosmology \citep{Brout19:sys}.  
A proper spectral model is also critical to distinguish between intrinsic color variations and non-Milky-Way-like dust \citep{Brout20, Thorp21}, the largest astrophysical systematic uncertainty for SN~Ia cosmology.

Although SN Ia distances have traditionally used optical light curves, recent work has revealed the promise of near-IR observations to improve both statistical and systematic uncertainties. 
Theory and small data sets both show that SNe Ia are more standard in the near-IR than the optical, and the effect of dust extinction is strongly mitigated \citep{Mandel11, Dhawan18, galbany2022}.

The most sophisticated algorithms for measuring SN distances currently do not have any true spectral model in the NIR.  
The most recent iteration of the popular SALT spectral model, SALT3 \citep{kenworthy2021}, did not attempt to have their model extend beyond 1~$\mu$m because of the lack of NIR data.  
BayeSN, a new hierarchical Bayesian model for time-dependent SN~Ia SEDs \citep{Mandel20}, was able to extend to the NIR, but only trained on photometry and not individual spectra.  
In the most robust effort yet in extending optical SN spectral models to the NIR, \citet{Pierel22} used the full public sample of appropriate SNe~Ia with any NIR data (photometry or spectra).  
The sample contained 166 SNe~Ia with NIR data, but only $\sim 50$ spectra with coverage beyond $\sim 1$~$\mu$m. 
The resulting model is useful for simulations of upcoming surveys \citep[e.g.,][]{Rose21} and general light-curve fitting, but is still insufficient for cosmological inference in the NIR.

Despite their unique utilities, IR spectra remain rare for transients.
In comparison to the $\sim 50,000$ optical transient spectra publicly available, fewer than 1000 NIR spectra of transients have been published.
By far, the largest source of NIR spectra of SNe is the Carnegie Supernova Project~II \citep[CSP~II;][]{hsiao2019}, which ran between 2011 and 2015, primarily using the FIRE spectrograph on the 6.5~m Magellan telescope. 
CSP~II obtained 909 NIR spectra of 249 unique SNe, a fraction of which is now publicly available.\footnote{\url{https://csp.obs.carnegiescience.edu/data}.}
The spectra were released with the publication of three sample papers: \citet{davis2019} focusing on SNe~II, \citet{shahbandeh2022} on stripped-envelope (SE) SNe, and \citet{lu2023} on normal SNe~Ia. 
Among these three papers, 495 spectra from 162 SNe were made public. 
The SN~II sample excluded all nebular spectra taken $> 300$ days post-explosion \citep{davis2019}. 
The SESN sample excluded a few objects with no clear optical classification or photometry near peak brightness \citep{shahbandeh2022}.
The SN~Ia sample excluded peculiar subclasses like SN\,2002cx-like (Iax), 2002ic-like, 2003fg-like (super-Chandrasekhar, SC), and Ca-strong objects.
(One SN~Ia-SC, LSQ14fmg, was published as a single-object paper; \citealp{hsiao2020}).
It also excluded objects with no photometric coverage around peak brightness in the $B$ band, and spectra later than 100~days from peak. 
For homogeneity, this sample only included spectra taken by FIRE in the low-resolution ($R\approx 400$) prism mode and not the high-resolution ($R\approx 6000$) echellette mode or spectra taken by other instruments.
With these cuts, only 54\% of the obtained spectra of SNe~Ia were made public \citep{lu2023}. 
CSP~II public data releases thus far had not contained any interacting SNe and other classes of transients.
Lastly, given the time period over which it operated, a significant portion (30--60\%, depending on the SN type) of CSP~II SNe still came from targeted surveys, which are biased towards luminous, massive galaxies.

The advancement in time-domain astronomy since 2015 has necessitated another NIR spectroscopic survey of astrophysical transients. 
Virtually all transients now are (or would have been) discovered by one of the many aforementioned untargeted optical surveys, and as such, our sample of nearby transients ($z \lesssim 0.01$) is not affected by the selection bias towards transients in bright, massive galaxies.
In addition, since 2015, many new classes of transients have been found or become well established, including TDEs \citep[see, e.g.,][]{arcavi2014, french2020,gezari2021,yao2023} and hydrogen/helium-poor interacting Type Icn SNe \citep{galyam2022, pellegrino2022, perley2022, davis2023}.
Their NIR evolution is either poorly observed or entirely unknown.
Finally, the analyses of incoming data from \textit{JWST} \citep[e.g.,][]{kwok2023,siebert2023, derkacy2023} and the ongoing planning efforts for \textit{Roman} require a robust and unbiased template of NIR spectra of transients of all types from the ground.
Below, we describe such a survey.

The Keck Infrared Transient Survey (KITS) is a NASA Key Strategic Mission Support (KSMS) program, which ran from 2022 February to 2023 July.
We provide the first data release, containing all NIR spectra obtained in the first observing semester between 2022 February and 2022 July. 
In Section~\ref{sec:strategy}, we describe the survey strategy and our target-selection criteria.
Sections~\ref{sec:observations} and \ref{sec:reduction}  describe our observational procedures and data reduction to ensure the reproducibility of our data products. 
The observed sample and the data included in this data release is described in Sections~\ref{sec:obs_sample} and ~\ref{sec:data}.
We provide a summary in Section~\ref{sec:summary}.

\begin{figure*}
    \centering
\textbf{Full KITS Sample} \\
\vspace{12pt}
\begin{minipage}[c]{0.49\textwidth}
\centering
    \textbf{By Unique Objects}
    \vspace{-1cm}
    \includegraphics[width=\textwidth]{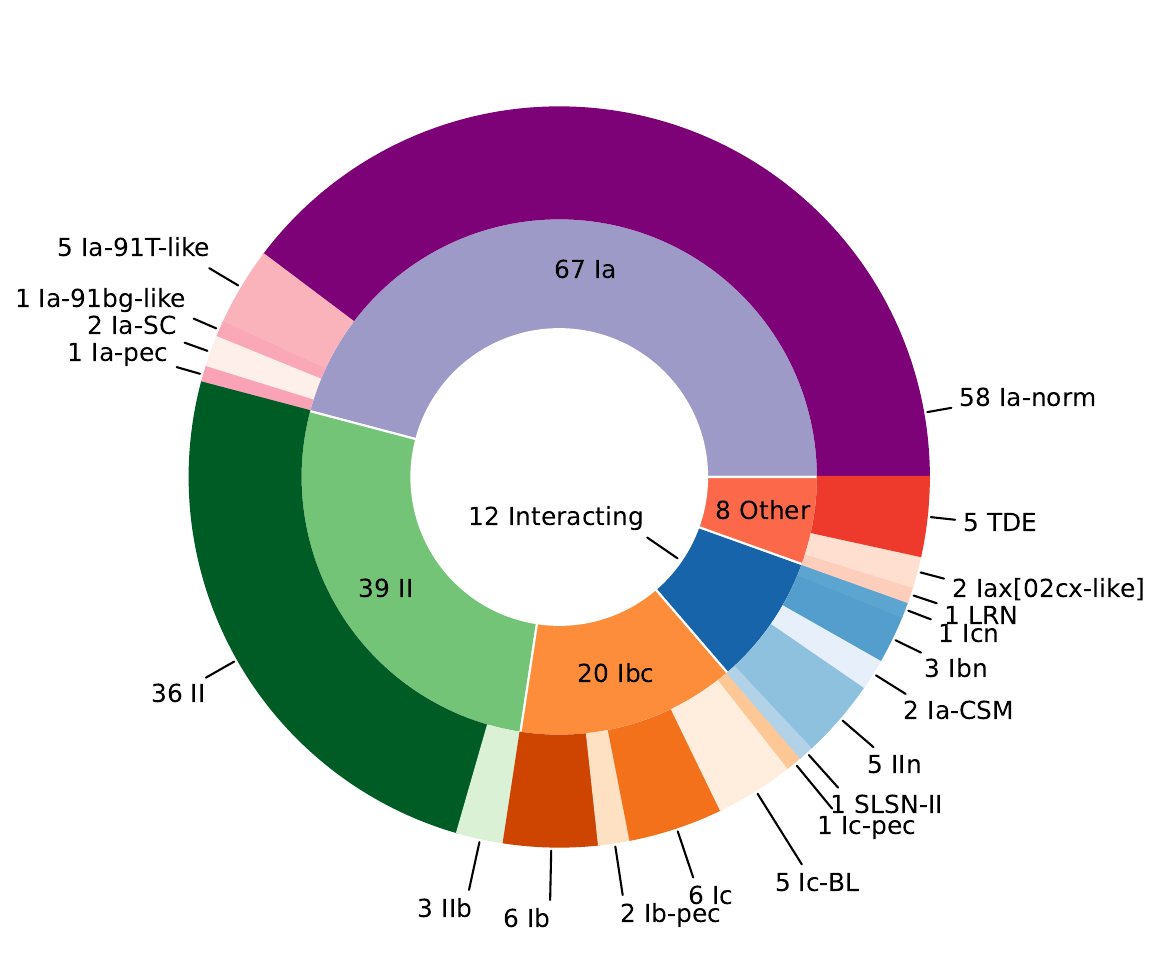}
\end{minipage}
\begin{minipage}[c]{0.49\textwidth}
\centering
    \textbf{By Spectra}
    \vspace{-1cm}
    \includegraphics[width=\textwidth]{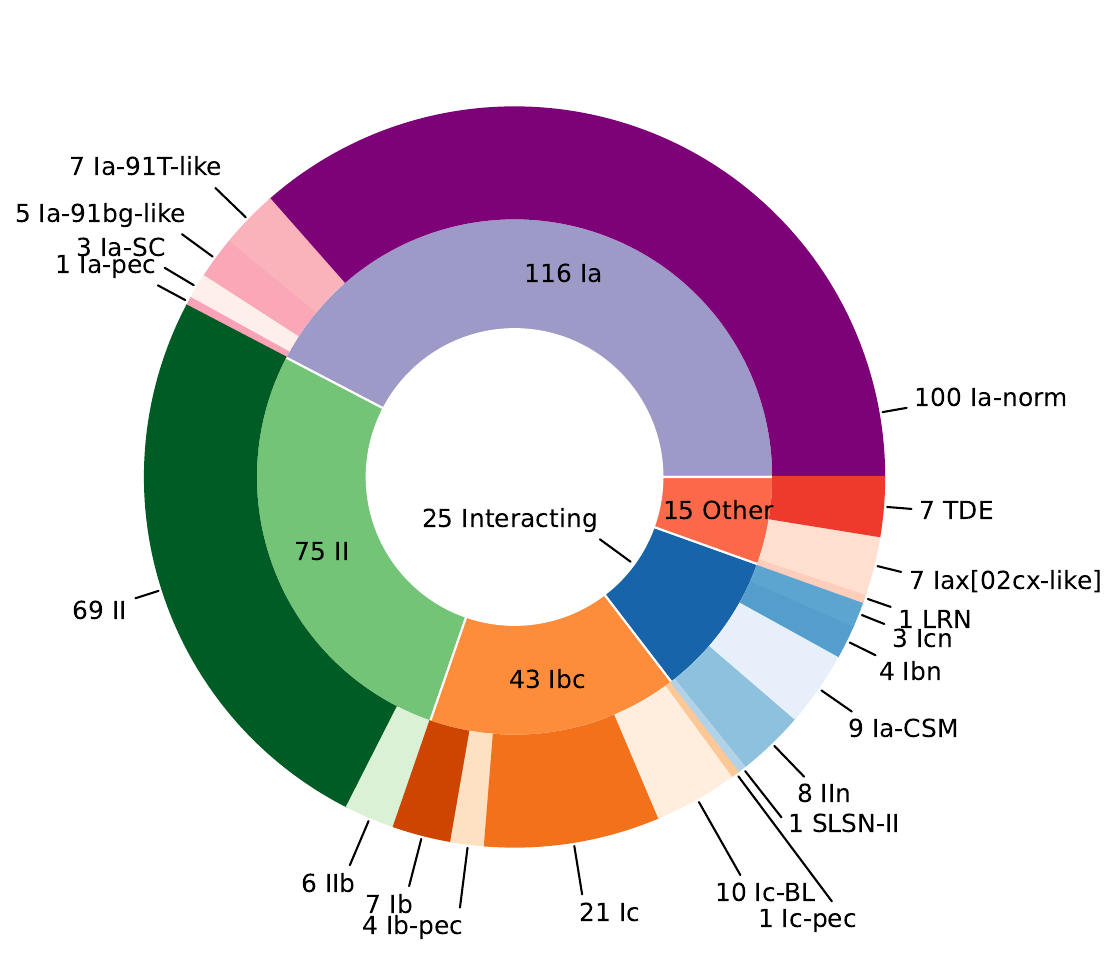} 
\end{minipage}
\textbf{This Data Release} \\
\hspace{0.03\textwidth}
\includegraphics[width=0.45\textwidth]{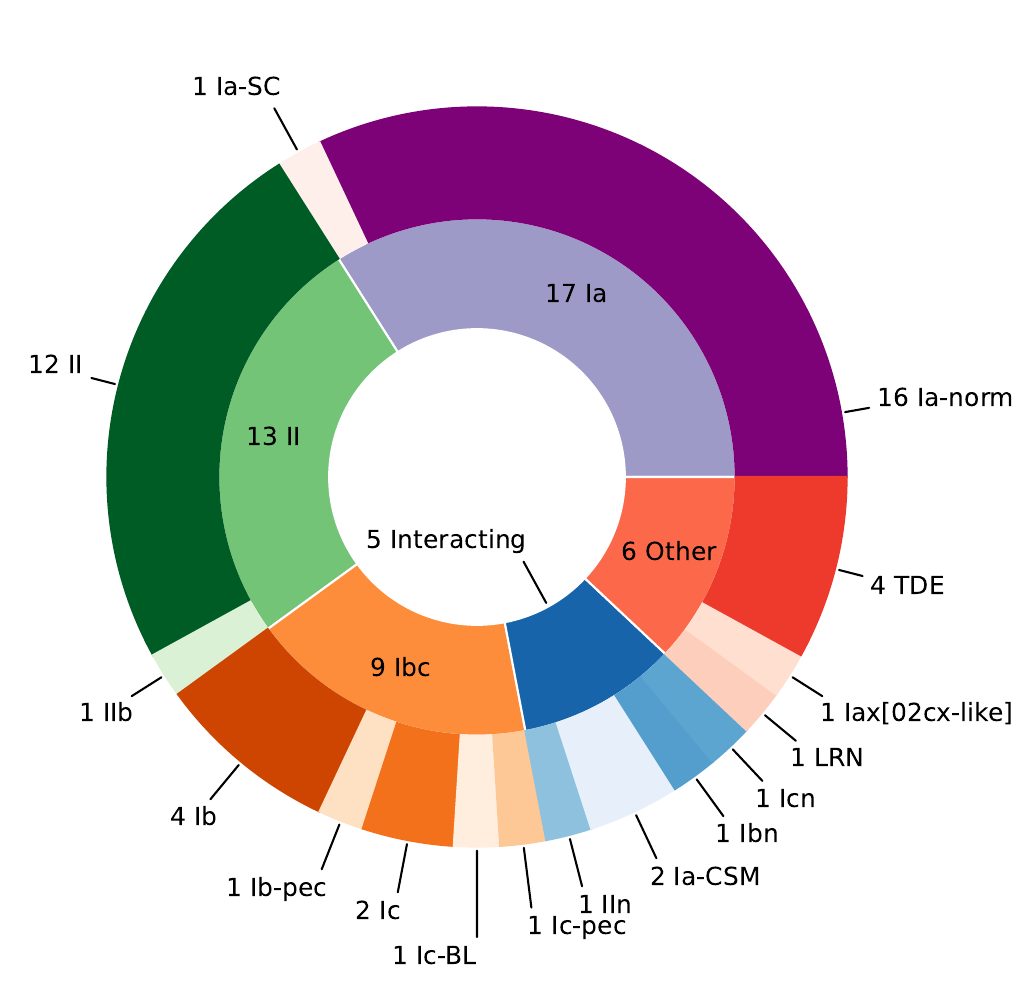}
\hfill
\includegraphics[width=0.48\textwidth]{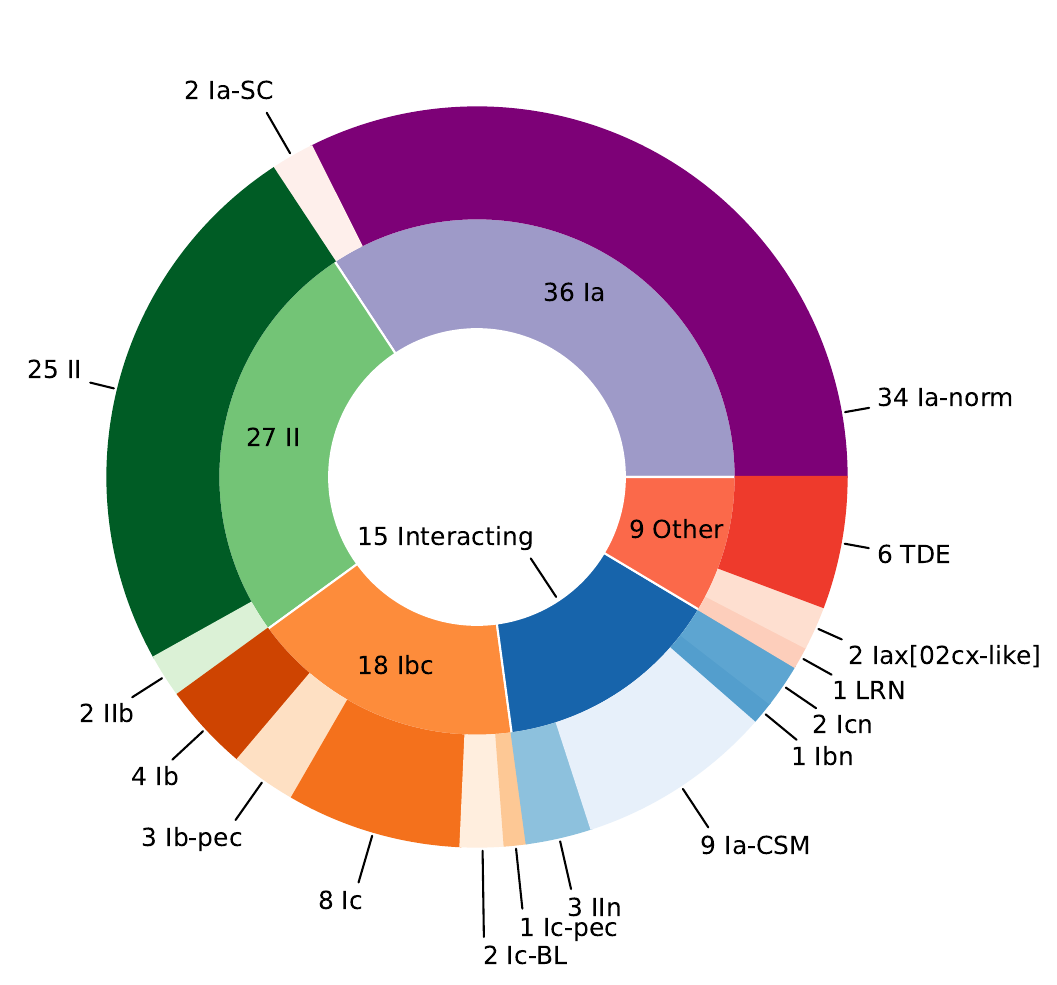}
    \caption{Pie charts showing numbers of SN types observed with KITS by unique events \textit{(left)} and individual spectra \textit{(light)}. 
    The \textit{top} row includes the entire KITS sample (274 spectra from 146 objects) while the \textit{bottom} row includes objects and spectra from this data release (105 spectra from 50 objects). 
    }
    \label{fig:type_pie_chart}
\end{figure*}

\section{Survey Strategy}\label{sec:strategy}
The overarching goal of KITS is to provide a large, publicly accessible NIR spectroscopic treasury of all types of astrophysical transients.  This program is in direct support of the mission success criteria of both {\it Roman} and {\it JWST}.

A key science driver of \textit{Roman} is to measure the expansion history of the Universe through luminosity-distance measurements. 
\textit{Roman}'s High-Latitude Time Domain Survey is designed to discover and measure distances to thousands of SNe~Ia \citep{Hounsell18, Rose21}.  
For $z<1$, {\it Roman} will observe the rest-frame NIR.  
Our distance-measurement techniques require a training set containing precise SED surfaces that span the wavelength, phase, and light-curve parameters of the the full SN~Ia parameter space.  
The lack of SN~Ia NIR spectra previously prevented accurate spectral models at these wavelengths \citep{Pierel22}.  
KITS addresses this issue by providing the necessary spectra to be incorporated into the models through the SALTShaker training process outlined in \citet{kenworthy2021}.
This will allow us to better plan for \textit{Roman} and better leverage its data.

Major goals for {\it JWST} are to detect the most distant and luminous objects in the Universe, determine when the first stars were born, and constrain the timing of the epoch of reionization \citep{gardner2006}.  
SNe, and in particular exotic, luminous SNe, may be the most-distant objects that will be discovered.  
Since these objects can occur only a few Myr after the first stars are formed, they can shine brighter than their nascent host galaxies and potentially be discovered at higher redshifts than galaxies.  
These explosions will also provide large amounts of ionizing photons, and measuring the rates and energetics of these SNe will determine their contribution to reionization.  
Even at $z \approx 20$, MIRI observes the rest-frame NIR, and accurate optical/NIR SEDs are necessary to identify and characterize high-redshift SNe that are too faint for spectroscopy.

Additionally, GO programs for both missions will undoubtedly study all other classes of transients, and NIR spectra will be critical to their understanding.  We planned KITS to obtain NIR spectra of rare events and of common events in epochs at which NIR spectra are rare.
These data will serve as comparison templates for new NIR observations of any transients from the ground or space.
They will also help with photometric classification of high-redshift transients that {\it Roman} or \textit{JWST} may serendipitously detect in the future.

In order to accomplish these goals, KITS focuses on three samples: (a) a flux-limited sample containing all transients brighter than 17~mag in a red optical band (usually ZTF $r$ or ATLAS $o$); (b) a volume-limited sample of all transients with $z < 0.01$; (c) and SNe~Ia with light-curve parameters or phases that are poorly sampled in the NIR.
Lastly, we aim to observe rare transients with little to no prior NIR spectroscopy. 
We have used our target-of-opportunity (ToO) observations to further obtain NIR spectra of very young transients.
Throughout the survey, we kept track of our progress for each subsurvey and adjusted our strategy to maximize the completeness and potential scientific output of our data, using \ysepz, our in-house open-source target and observation management system \citep{coulter2022, coulter2023}.
We select transients discovered by the aforementioned public transient surveys.

\section{Observations}\label{sec:observations}
KITS operated over about 12 half nights per semester in the 2022A (2022 Feb--Jul), 2022B (2022 Aug--2023 Jan), and 2023A (2023 Feb--Jul) semesters, with observations occurring roughly once every two weeks.
All observations in 2022A and 2022B were in the second half of the night, while 2023A observations were in the first half. 
The last night of 2023A (2023 July 29 UT) was the only full-night observation. 
We obtained four ToO spectra with KITS to observe objects at crucial phases (e.g., young objects, or SNe~Ia at previously unobserved phases).
Table~\ref{tab:obs_log} summarizes all KITS nights and Table~\ref{tab:KITS_SNe} lists all SNe observed by KITS in the 2022A semester.

All spectra (except one) were obtained with the Near-InfraRed Echellette Spectrometer (NIRES) on the Keck~II telescope. 
NIRES is the latest member of the TripleSpec family of four NIR spectrographs. 
\citet{wilson2004} provides an overview of this design. 
To make the same spectrograph optics work on different telescopes with different focal ratios, fore optics are installed to convert the incoming beam from the telescope to a uniform $f$/10.7.
With Keck's large aperture, the field of view decreases proportionally.
NIRES's single slit is $0.55'' \times 18''$, and the slit-viewing camera has a field of view (FoV) of only $1.8' \times 1.8'.$\footnote{\url{https://www2.keck.hawaii.edu/inst/nires/}}
As a result of the small slit-viewing camera FoV, NIRES has an off-axis optical guide camera to help ensure that there is a star on which to guide.
Crucially, the position angle (PA) of NIRES has to be selected such that there is a bright guide star in the FoV of the guide camera.

We follow a standard procedure for the observations to ensure the uniformity of data quality, which we outline below.
For each night of observation, collaboration members query the database on \ysepz\ for all transients that fall into our aforementioned subsamples, and request observations.
The observer in charge downloads all observation requests from \ysepz\ and runs a number of \texttt{Python} scripts to create finder charts with offset stars, to select a PA with a guide star in the guider, and to search for a nearby A0~V star necessary to correct for telluric absorption. 
The script also computes the rising/setting time of each target (taking into account Keck~II's western pointing limit owing to the Nasmyth platform), suggested exposure times, Moon distance, and the telescope azimuth wrap in which the target is observable. 
We schedule the night using these outputs, which help us maximize the observing efficiency. 
These scripts and their documentation are publicly available for other NIRES users\footnote{\url{https://github.com/stinyanont/YSE_finder}}.

\begin{table}[h]
    \caption{KITS Observation Log}
    \begin{tabularx}{0.98\linewidth}{llll}
    \toprule
        UT Date & Night Half & Hours available & Targets \\ \hline
2022-02-13 & second & 5.5 & 9 \\
2022-02-22 & second & 5.4 & 9 \\
2022-03-11 & second & 5.2 & 8 \\
2022-03-24 & second & 5.1 & 10 \\
2022-04-15 & second & 4.9 & 9 \\
2022-04-22 & second & 4.8 & 0 \\
2022-05-09 & second & 4.7 & 8 \\
2022-05-17 & MOSFIRE ToO    & 1.0   & 1 \\
2022-05-20 & second & 4.6 & 0 \\
2022-06-07 & second & 4.5 & 10 \\
2022-06-11 & ToO & 1.0 & 2 \\
2022-06-22 & second & 4.4 & 7 \\
2022-07-09 & second & 4.5 & 8 \\
2022-07-19 & second & 4.5 & 11 \\
2022-08-05 & second & 4.7 & 9 \\
2022-08-18 & second & 4.8 & 9 \\
2022-09-09 & second & 5.0 & 8 \\
2022-09-18 & second & 5.1 & 8 \\
2022-10-06 & second & 5.3 & 10 \\
2022-10-15 & second & 5.4 & 7 \\
2022-11-06 & second & 5.5 & 8 \\
2022-11-17 & second & 5.6 & 12 \\
2022-11-18 & ToO & 1.0 & 2 \\
2022-12-01 & second & 5.6 & 3 \\
2022-12-15 & second & 5.7 & 6 \\
2022-12-31 & second & 5.7 & 14 \\
2023-02-01 & first & 5.5 & 0 \\
2023-02-08 & first & 5.5 & 8 \\
2023-02-27 & first & 5.3 & 0 \\
2023-03-12 & first & 5.2 & 0 \\
2023-03-30 & first & 5.1 & 9 \\
2023-04-07 & ToO & 1.0 & 1 \\
2023-04-10 & first & 4.9 & 10 \\
2023-04-27 & first & 4.8 & 5 \\
2023-05-05 & first & 4.7 & 11 \\
2023-05-28 & first & 4.5 & 11 \\
2023-06-07 & first & 4.5 & 7 \\
2023-07-05 & first & 4.5 & 7 \\
2023-07-29 & full & 9.2 & 17 \\ \hline
    \end{tabularx}
    \label{tab:obs_log}
\end{table}

For each observing night, we obtain flat-field images using the dome-flat lamp.
We find that using ten standard flat exposures, 120~s each, is sufficient.
Flats with the lamp off (``dark frames") are unnecessary because at this exposure time, the lamp-off flats have significant flux in the $K$ band, and do not capture the dark current. 
Observations of comparison lamps are also not necessary (but always taken) because we use IR night-sky lines to perform wavelength calibrations. 

At the beginning of the night, the operator runs the MIRA software to focus the telescope.
To acquire each target, we first take a pair of images of the target field with the slit-viewing camera.  
We are developing a pipeline that can automatically reduce NIRES slit-viewing camera images and use them to measure photometry.
This will be included in the next data release.
We identify the target and offset the telescope to place the target in the slit.
Another image is taken for confirmation, and then the spectroscopic sequence is started, using exposure times suggested by our observation preparation script. 
All science observations are performed in an ABBA dithering pattern with the A and B positions $6''$ apart on the slit.
This observing strategy allows us to subtract the bright NIR sky lines.
For some observations, we perform two cycles of ABBA or simply an additional AB pair, with the determining factors being the total exposure time and the restriction that individual exposures be at most 300~s due to the saturating sky lines.
Immediately before or after each science observation, we also obtain a spectrum of an A0~V star for flux and telluric calibration.
These stars are also observed with an ABBA pattern, but with a $10''$ offset to prevent persistence from observing a bright star to interfere with subsequent science observations.

One ToO spectrum of SN\,2022jzc on 2022 May 17 was obtained using the Multi-Object Spectrometer for Infra-Red Exploration \citep[MOSFIRE;][]{mclean2012} in the long-slit configuration with a $0.7''$ slit width. 
With MOSFIRE, spectra are obtained one filter at a time, and within the 1~hr interrupt we only had sufficient time to observed in the $Y$, $J$, and $K$ bands. 
A similar calibration procedure was followed for the MOSFIRE observation.

In parallel to Keck/NIRES observations, we also obtain observations of very bright objects using the SpeX spectrograph \citep{rayner2003} on the 3~m NASA InfraRed Telescope Facility (IRTF), and the TripleSpec spectrograph on the 4.1~m Southern Astrophysical Research (SOAR) Telescope.
These observations are not included in the discussion of KITS in this paper, but will appear in the next and final data release. 

\begin{deluxetable*}{lrrllllcl}
\tabletypesize{\scriptsize}
\tablecaption{KITS 2022A Transients\label{tab:KITS_SNe}}
\tablehead{\colhead{AT/SN} & \colhead{RA} (J2000) & \colhead{Dec} (J2000) & \colhead{Spec. Type} & \colhead{Redshift} & \colhead{Ref MJD} & \colhead{Ref MJD Type}&\colhead{$N_{\rm obs}$} & \colhead{Reference} 
}
\startdata
2020ohl & 17:03:36.5 &  +62:01:32.34 & TDE       &    0.01671    & 59014.4	& explosion  &  1  &    \protect\cite{hinkle2020,hinkle2022} \\
2021biy & 12:42:04.0 &  +32:32:07.87 & LRN       &    0.002021   & 59244.5	& explosion  &  1  &    \protect\cite{cai2022} \\
2022fw  & 12:23:54.0 &  -03:26:37.88 & SN Ia     &    0.0067     & 59601.9	& peak  &  4  &    \protect\cite{hosseinzadeh2022fw} \\
2022jo  & 13:00:37.7 &  +28:03:25.76 & SN II     &    0.0265     & 59581.5	& explosion  &  1  &    \protect\cite{li2022jo} \\
2022mm  & 11:58:25.1 &  -14:31:11.50 & SN II     &    0.013      & 59590.0	& explosion  &  1  &    \protect\cite{reguitti2022mm} \\
2022abq & 13:22:56.8 &  +28:19:08.87 & SN II     &    0.007979   & 59599.5	& explosion  &  2  &    \protect\cite{ochner2022} \\
2022afc & 07:56:45.0 &  +26:53:07.36 & SN Ib     &    0.028      & 59621.0	& peak  &  1  &     \protect\cite{davis2022afc}\\
2022ann & 10:17:29.7 &  -02:25:35.44 & SN Icn    &    0.049      & 59609.6	& peak  &  2  &    \protect\cite{davis2023} \\
2022baw & 12:48:50.2 &  +37:14:51.76 & SN Ia     &    0.04       & 59622.6	& peak  &  1  &     \protect\cite{tucker2022atel}\\
2022bck & 13:24:35.1 &  -20:11:07.48 & SN Ib     &    0.026      & 59627.8	& peak  &  1  &    \protect\cite{lyman2022tns} \\
2022bdu & 09:36:52.2 &  +37:41:38.99 & SN Ic     &    0.015      & 59628.0	& peak  &  2  &    \protect\cite{tucker2022atel} \\
2022bdw & 08:25:10.4 &  +18:34:57.50 & TDE       &    0.03782    & 59632.0	& peak  &  1  &    \protect\cite{arcavi2022} \\
2022bse & 07:01:02.3 &  +51:15:55.68 & SN II     &    0.020954   & 59618.5	& explosion  &  1  &    \protect\cite{srivastav2022} \\
2022crr & 15:24:49.1 &  -21:23:21.73 & SN Ic-BL  &    0.0188     & 59636.1	& peak  &  2  &    \protect\cite{davis2022crr} \\
2022crv & 09:54:25.9 &  -25:42:11.16 & SN Ib     &    0.008091   & 59647.0	& peak  &  1  &    \protect\cite{andrews2022crv} \\
2022cvr & 14:01:21.6 &  +37:18:56.97 & SN Ia     &    0.064      & 59640.9	& peak  &  1  &    \protect\cite{ztf2022cvr} \\
2022dbl & 12:20:45.0 &  +49:33:04.68 & TDE       &    0.0284     & 59639.3	& peak  &  2  &     \protect\cite{arcavi2022dbl} \\
2022dml & 16:17:29.1 &  +14:25:04.61 & SN II     &    0.03       & 59635.0	& explosion  &  1  &    \protect\cite{taggart2022dml, burke2022dml} \\
2022dsb & 15:42:21.7 &  -22:40:14.04 & TDE       &    0.023      & 59630.0	& explosion  &  2  &     \protect\cite{fulton2022dsb} \\
2022dtv & 14:36:35.2 &  +11:56:21.40 & SN Ia     &    0.028617   & 59654.0	& peak  &  2  &    \protect\cite{hinds2022dtv} \\
2022eat & 11:11:31.0 &  +19:49:38.17 & SN Ia     &    0.027      & 59657.8	& peak  &  2  &    \protect\cite{chu2022eat} \\
2022erq & 18:33:25.4 &  +44:05:11.65 & SN Ia-CSM &    0.066      & 59682.0	& peak  &  6  &    \protect\cite{li2022erq} \\
2022erw & 10:50:57.8 &  -02:08:59.28 & SN Ia     &    0.015      & 59664.0	& peak  &  1  &     \protect\cite{moore2022erw} \\
2022esa & 16:53:57.6 &  -09:42:10.26 & SN Ia-CSM &    0.023      & 59709.5	& peak  &  3  &    \protect\cite{lu2022esa} \\
2022ewj & 10:46:34.6 &  +13:45:17.03 & SN II     &    0.010134   & 59655.0	& explosion  &  1  &    \protect\cite{tagchi2022ewj} \\
2022exc & 17:15:02.1 &  +60:12:58.79 & SN Ia     &    0.020123   & 59671.1	& peak  &  5  &    \protect\cite{do2022exc} \\
2022eyj & 11:18:00.6 &  +07:50:44.66 & SN Ia     &    0.021103   & 59662.5	& peak  &  1  &     \protect\cite{balcon2022eyj} \\
2022eyw & 12:43:60.0 &  +62:19:48.29 & SN Iax    &    0.009      & 59678.8	& peak  &  2  &    \protect\cite{tagchi2022eyw} \\
2022fcc & 14:15:54.8 &  +03:36:14.60 & SN Ia     &    0.025851   & 59682.0	& peak  &  1  &    \protect\cite{pellegrino2022fcc} \\
2022frl & 15:21:33.1 &  -07:26:52.04 & SN Ib-pec &    0.006      & 59691.8	& peak  &  3  &    \protect\cite{tucker2022frl} \\
2022frn & 12:59:51.8 &  +27:56:36.66 & SN Ia     &    0.023      & 59684.1	& peak  &  2  &    \protect\cite{balcon2022frn, fulton2022frn} \\
2022hrs & 12:43:34.3 &  +11:34:35.87 & SN Ia     &    0.0047     & 59698.5	& peak  &  3  &    \protect\cite{balcon2022hrs} \\
2022hsu & 22:11:37.7 &  +46:18:40.03 & SN IIn    &    0.018      & 59710.8	& peak  &  3  &    \protect\cite{ashall2022hsu, taggart2022hsu} \\
2022ihx & 19:16:38.4 &  +61:41:15.48 & SN Ibn    &    0.033      & 59700.1	& peak  &  1  &     \protect\cite{pellegrino2022ihx} \\
2022iid & 18:15:38.7 &  +73:08:06.05 & SN II     &    0.014      & 59701.5	& explosion  &  3  &    \protect\cite{fulton2022iid} \\
2022ilv & 15:10:44.3 &  -11:35:57.99 & SN Ia-SC  &    0.031      & 59707.2	& peak  &  2  &    \protect\cite{srivastav2023} \\
2022jli & 00:34:45.7 &  -08:23:12.16 & SN Ic     &    0.006      & 59704.2	& discovery  &  6  &    \protect\cite{monard2022, grzegorzek2022jli} \\
2022joj & 14:41:40.1 &  +03:00:24.33 & SN Ia     &    0.03       & 59724.1	& peak  &  1  &     \protect\cite{newsome2022joj} \\
2022jzc & 12:05:28.67 & +50:31:36.80 & SN II     &    0.0029     & 59714.3  & explosion & 1 &  \protect\cite{bruch2022jzc} \\
2022kla & 16:44:33.2 &  +38:55:03.25 & SN Ia     &    0.037112   & 59734.8	& peak  &  2  &    \protect\cite{ztf2022kla} \\
2022lxg & 19:15:23.6 &  +48:19:27.70 & SN II     &    0.0214     & 59731.5	& explosion  &  4  &    \protect\cite{ashall2022lxg}, this work \\
2022mji & 09:42:54.1 &  +31:51:03.67 & SN II     &    0.004      & 59731.9	& explosion  &  1  &    \protect\cite{sollerman2022mji} \\
2022mxv & 23:51:05.1 &  +20:09:08.96 & SN II     &    0.014046   & 59745.5	& explosion  &  6  &     \protect\cite{davis2022mxv} \\
2022mya & 17:21:08.1 &  +16:03:32.47 & SN Ib     &    0.03       & 59759.1	& peak  &  1  &    \protect\cite{sollerman2022mya} \\
2022nag & 18:05:00.7 &  +09:28:47.86 & SN Ia     &    0.020954   & 59757.1	& peak  &  3  &    \protect\cite{ztf2022nag} \\
2022ngb & 18:56:51.5 &  +36:37:07.82 & SN IIb    &    0.009      & 59777.6	& peak  &  2  &    \protect\cite{izzo2022ngb} \\
2022ojo & 01:44:35.6 &  +37:41:50.72 & SN II     &    0.019      & 59755.5	& explosion  &  3  &    \protect\cite{desai2022ojo} \\
2022oqm & 15:09:08.2 &  +52:32:05.14 & SN Ic-pec &    0.012      & 59783.9	& peak  &  1  &    \protect\cite{irani2022} \\
2022osg & 20:29:49.0 &  -02:01:41.11 & SN Ia     &    0.01858    & 59785.1	& peak  &  2  &    \protect\cite{lidman2022osg} \\
2022ovq & 02:01:59.9 &  +21:06:23.45 & SN Ia     &    0.030298   & 59786.4	& peak  &  3  &    \protect\cite{hinkle2022ovq} \\
\enddata
\end{deluxetable*}

\section{Data Reduction}\label{sec:reduction}
The data are typically downloaded the following day and we process them for quick quality assessment using \texttt{IDL} software \texttt{spextool} \citep{cushing2004} and \texttt{xtellcor} \citep{vacca2003}, which are standard in NIR spectroscopy. 
While widely used and well tested, \texttt{spextool} and \texttt{xtellcor} require intensive user input to extract data and perform telluric correction.
They also require a paid \texttt{IDL} license to run.
In order to perform a uniform data reduction for this public data release, and make our process as reproducible as possible, we uniformly reprocess our data using the much more autonomous \texttt{Python}-based open-source facility spectroscopic reduction software \texttt{Pypeit} \citep{prochaska2020, pypeit2020}.
We specifically use \texttt{v.1.13.0}, which has many useful functionalities to support NIRES observations.

There are three steps to process a night of data with \texttt{Pypeit}: preparation, spectral extraction, and calibrations.
We follow the \texttt{Pypeit} documentation, and wrote additional helper scripts to make the process more automated.
In the preparation step, \texttt{Pypeit} goes through the raw data directory and classifies files as science, standard, or different types of calibration. 
\texttt{Pypeit} can also figure out the dither pattern used in the observations and automatically assign a background frame for subtraction.
The script outputs a setup file summarizing the observing log for a user to verify that everything is correct. 
We remove dome flats with lamp off and wavelength comparison-lamp frames from this file, so they do not get processed further. 
We also add two flags to ask \texttt{Pypeit} to only attempt to find one source in each science and standard frame in the next step.

After the preparation is done and we have verified that everything in the setup file is correct, we can execute this file by running the \texttt{run\_pypeit} command. 
This takes about 2~hr on our computer node, but could take up to several hours on a personal computer. 
The script creates a master flat field using dome-flat observations.
Dome flats are also used to identify the illuminated area of the detector.
The script uses night-sky emission lines in science frames to perform wavelength calibration and measure the tilt of the spectra.
Standard observations automatically get assigned wavelength solution from the science observations closest in time and airmass.
It is crucial to note that \texttt{Pypeit} provides wavelengths in vacuum, and not in air as is typical for NIR lines. 
After all the calibrations are prepared, \texttt{Pypeit} goes through all science and standard exposures, applies these calibrations, automatically identifies a source, and performs spectral extraction. 
The resulting extracted spectra are stored in \texttt{spec1d} files, along with calibrated two-dimensional (2D) spectra in \texttt{spec2d} files. 

The last step is to calibrate the extracted spectra to go from extracted 1D spectra to flux-calibrated, telluric-corrected, science-ready spectra.
\texttt{Pypeit} provides commands to do each step separately, and some steps also need a separate parameter file. 
To speed up the process, we wrote a \texttt{Python} script to create a bash script containing all commands we need to perform all these calibrations to all our data, along with necessary parameter files. 
It also matches science targets to an appropriate telluric star.

\texttt{Pypeit} first computes the sensitivity function by comparing observations of each A0~V star to a model spectrum and the telluric absorption spectrum at Maunakea. 
This sensitivity function is then used to flux calibrate and coadd science observations. 
The coaddition step also stitches different orders of the spectrum together. 
We then fit the telluric model to the A0~V star observation using a telluric model grid specifically for Maunakea supplied by \texttt{Pypeit}.
In this step, we found that we need to add an optional parameter \texttt{polyorder = 5} to the default telluric parameter file to get a good fit. 
This is the order of a polynomial used to approximate the star's continuum across hydrogen absorption lines.

Finally, the telluric model is applied to the science observation to produce a final, science-ready spectrum. 
This last step is not currently supported by \texttt{Pypeit}. 
It can fit a telluric model to the science spectrum and apply the resulting model to the same spectrum.
For a generic target, this method normally uses a polynomial to approximate the science spectrum.
This method does not work for SN spectra, which often have spectral features coinciding with the telluric absorption lines, especially the $\rm CO_2$ lines around 2~$\mu$m coinciding with the He 2.0581~$\mu$m line. 
We have another custom \texttt{python} script to take the telluric model fitted to a standard-star observation and apply it to a corresponding science observation. 
However, this is the step in which further pipeline development can improve the resulting spectra quality.

To get from raw data to the final calibrated NIRES spectrum for each night, the user only needs to run three different commands and manually check two intermediate setup files. 
This workflow considerably reduces the workload and active time required to process a night of observation in comparison to running \texttt{spextool} (which usually takes an entire day to reduce data from one half night), and ensures that the resulting spectra are reproducible.

\section{Observed Sample}\label{sec:obs_sample}

\subsection{Observed Classes}

Figure~\ref{fig:type_pie_chart} displays pie charts conveying the fractions of the KITS sample that correspond to different transient classes.  
We present these fractions for distinct objects and observed spectra, both for the full KITS sample and what is included in the first data release (corresponding to semester 2022A).
In total, we obtained 274 spectra of 146 objects. 
This data release contains 105 spectra of 50 objects. 
In addition to the common SN classes, we observed a number of rare transients, including 2 SNe~Iax, 2 ``super-Chandrasekhar'' SNe~Ia, 4 TDEs, 1 luminous red nova (LRN), 3 SNe~Ibn, and 1 SN~Icn.
Data for all four TDEs (AT~2020ohl, 2022bdw, 2022dbl, and 2022dsb), the LRN (AT~2021biy), one SN~Iax (SN~2022eyw), and one SN~Ibn (SN~2022ihx) are included in this data release. 
Data for three objects --- SNe\,2022ann (Type Icn), 2022oqm (a peculiar, calcium-rich SN~Ic), and 2022joj (peculiar SN Ia) --- have already been published by \citet{davis2023}, \citet{karthik2023}, and \citet{padilla2023}, respectively, and for completeness we include those data as part of this data release. 
In addition, KITS observations from later semesters of SN\,2022pul have been used in conjunction with \textit{JWST} observations to study its peculiar nature \citep{siebert2023, kwok2023}. 

The release of the full sample is expected in mid-2024.
The full release will include spectra from IRTF/SpeX and SOAR/TripleSpec, further improving the NIR spectroscopic coverage of our sample. 
It will also include photometry of KITS objects from the slit-viewing camera of NIRES, and Gemini NIRI and FLAMINGOS-2.

\subsection{Completeness of the flux- and volume-limited samples}

\begin{figure}
    \centering
    \includegraphics[width=0.95\linewidth]{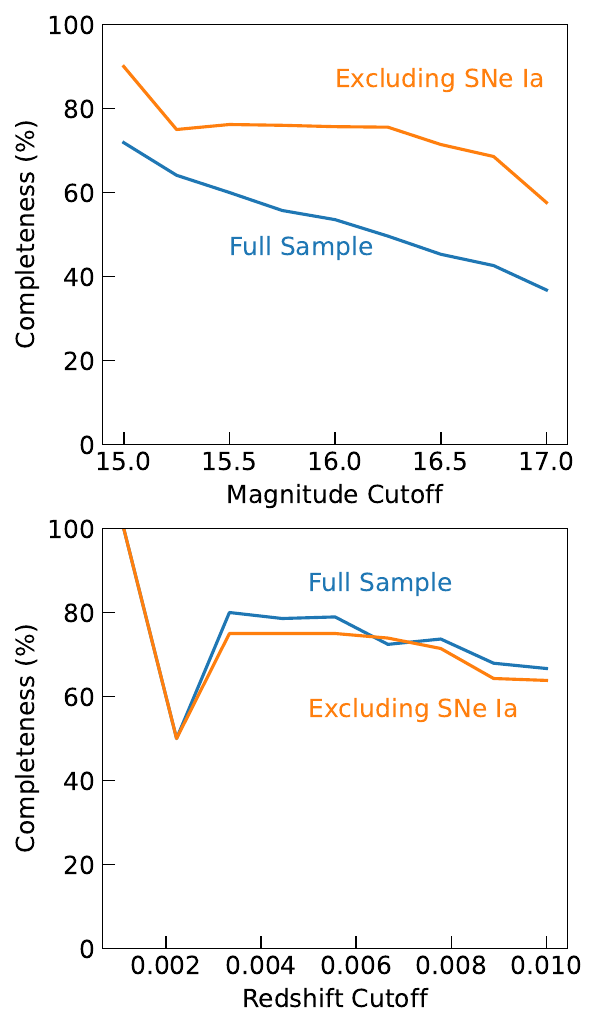}
    \caption{KITS completeness as a function of maximum magnitude in any reported visible bands, most commonly $r$ or ATLAS $o$ P(top) and maximum redshift (Bottom).  
    The completeness of the full sample is plotted in blue and that excluding SNe~Ia is plotted in orange. 
    }
    \label{fig:completeness}
\end{figure}

As described in Section~\ref{sec:strategy}, a main goal of KITS is to survey the full variety of transients without regard to class.  
Flux- and volume-limited samples are ideal ways to produce samples of rare objects with easily understood selection effects.  
A volume-limited sample targets low-luminosity transients that could be missed in a flux-limited survey.  
A flux-limited survey can access rare (and luminous) objects that may not occur in a relatively small volume.

We determine the full volume-limited sample by querying \ysepz\ for all transients with $z < 0.01$ that are discovered between $< 2$ weeks before our first night and on our last night.
For the flux-limited sample, we run a similar query but require a peak magnitude from any public survey of $< 17$~mag in the visible (typically either the ZTF $r$ band or ATLAS $o$ band).
While this selection should ideally have been done in the NIR, the present lack of such NIR facility for time-domain astronomy prevented us from doing so.
We went through every object on \ysepz\ to ensure that the photometric point used to determine the peak magnitude is not spurious. 
Objects with TNS classifications CV, VarStar, Nova, and ``Other'' are excluded.
We also exclude objects with $\delta < -30^\circ$, which are not easily visible from Keck.
Moreover, we exclude objects that are not visible during the half-nights allocated to KITS.
There are only 7 objects in the flux-limited sample and 1 objects in the volume-limited sample that fall into this category, so excluding them does not affect the completeness significantly. 
The fraction of these objects that has at least one NIRES spectrum is our completeness. 

Figure~\ref{fig:completeness} shows the completeness of our flux- and volume-limited samples.
The top panel shows the completeness as a function of a magnitude cutoff and the bottom panel indicates the completeness as a function of a redshift cutoff. 
Our volume ($z<0.01$) and flux ($r$/$o$ \textbf{}$< 17$~mag) limited samples are 37\% (103/280 objects observed) and 54\% (41/76 objects observed) complete, respectively.
The flux-limited sample is dominated by SNe~Ia.
Because of their homogeneity and the limited observing time, we did not observe the majority of SNe~Ia in the flux-limited sample near peak brightness.
The flux-limited sample, excluding SNe~Ia, is 58\% complete (53/92 objects observed).
The completeness fraction of the flux-limited sample, excluding SNe~Ia, improves to 71\% at 16.5~mag, while the volume-limited sample (including SNe~Ia) is 79\% complete at $z = 0.005$. 
Furthermore, bright objects missing from the flux-limited sample have been observed with IRTF/SpeX or SOAR/TripleSpec. 
Data from those observations, along with the improved statistics, will be available with the next data release.

\begin{figure*}
\centering
    \includegraphics[width=\textwidth]{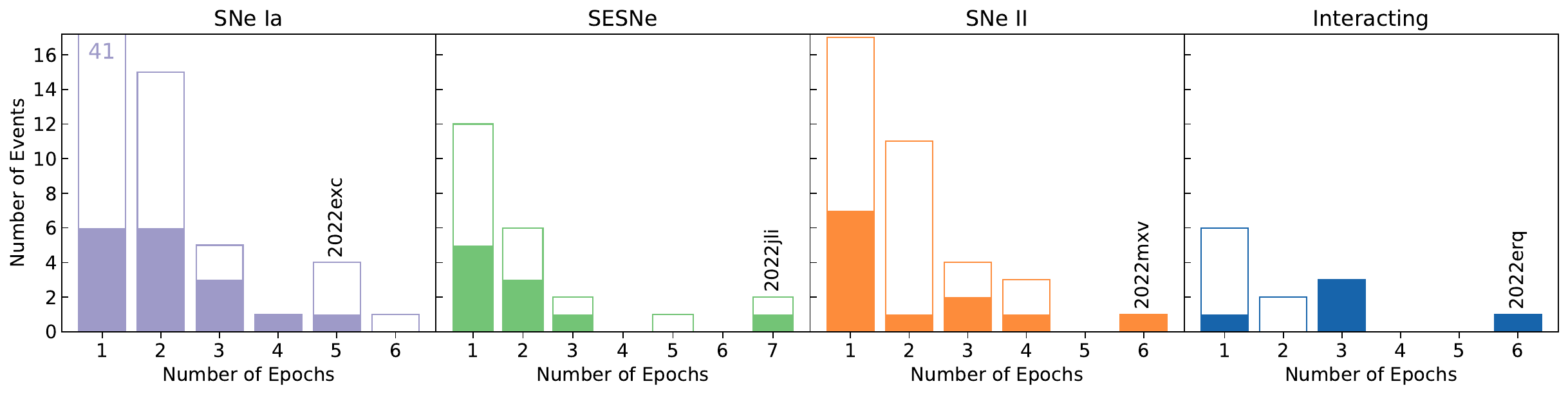}
    \caption{Histograms showing the distribution of number of KITS spectra per SN for different SN subtypes.
    The full sample is shown in unfilled bars; those included in this data release are filled. 
    The object in this data release with the most observations in each type is labeled in the plot.
    The \textbf{left} panel for SNe~Ia includes normal SNe~Ia, 91T-like, 91bg-like, and super-Chandra, while excluding SNe~Iax. 
    The \textbf{middle-left} panel for SESNe includes SNe~IIb, Ib, Ic, and Ic-BL, along with objects labeled peculiar. 
    The \textbf{middle-right} panel includes all SNe~II that are not IIb or IIn. 
    Finally, the \textbf{right} panel includes SNe~Ia-CSM, Ibn, Icn, and IIn.
    We attempt to obtain multiple spectra for every object, but in many cases it was not possible owing to (for example) the visibility constraint, the source fading, or scheduling issues. }
    \label{fig:epoch_num_distribution}
\end{figure*}

\begin{figure*}
    \centering
    \includegraphics[width=\textwidth]{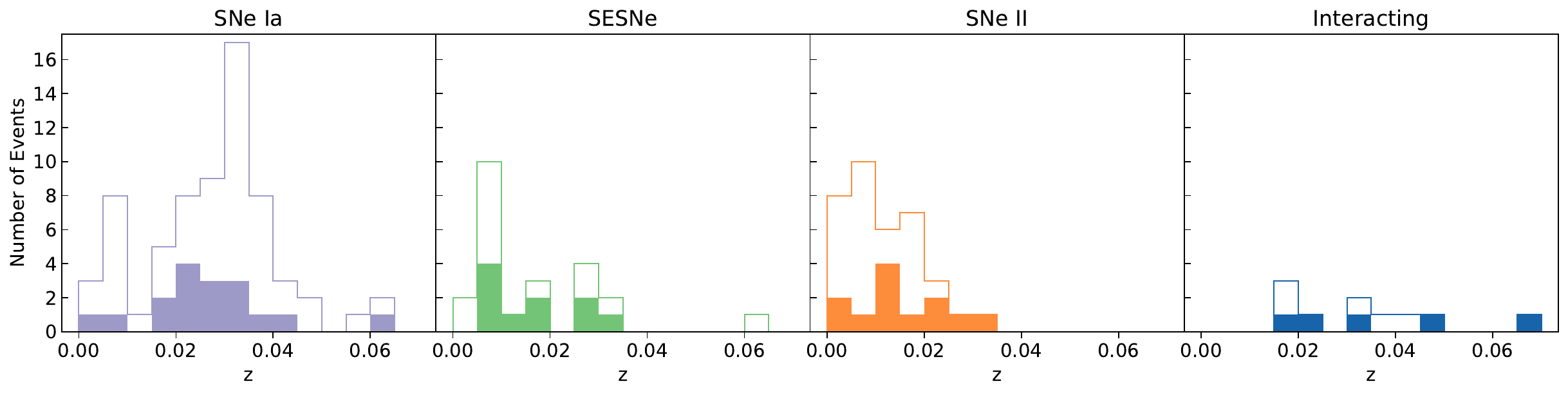}
    \caption{Histograms showing the distribution of redshift of SNe observed by KITS, with SNe~Ia, SESNe, SNe~II, and interacting objects plotted from left to right (same subtypes as plotted in Fig.~\ref{fig:epoch_num_distribution}).
    The full sample is shown in unfilled bars; objects included in this data release are filled. 
    }
    \label{fig:redshift_distribution}
\end{figure*}

\begin{figure*}
    \includegraphics[width=\linewidth]{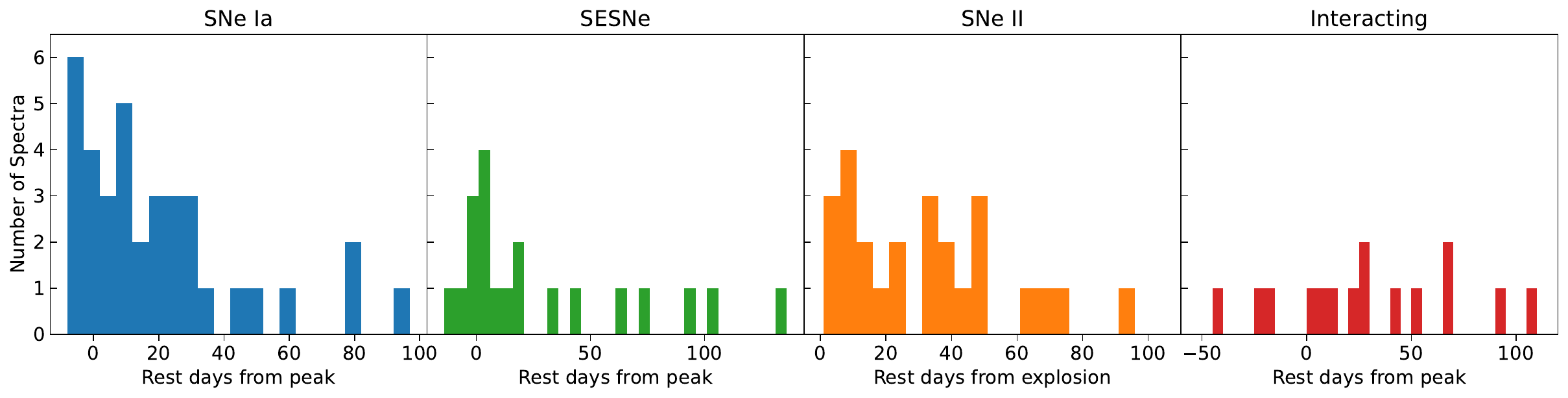}
    \caption{Histograms showing the distribution of phases of SN spectra observed by KITS, with SNe~Ia, SESNe, SNe~II, and interacting objects plotted from left to right. 
    These only include spectra from this data release.
    }
    \label{fig:phase_distribution}
\end{figure*}

\subsection{Sample Statistics}
Figure~\ref{fig:epoch_num_distribution} illustrates histograms of the number of observations per event for each common SN subtype. 
The full sample is shown, with the subset included in this data release indicated in shaded area. 
From left to right, we display histograms for SNe~Ia, including all subtypes except the interacting SNe~Ia-CSM; SESNe including Types IIb, Ib, Ic, and Ic-BL, along with peculiar hydrogen-poor events; SNe~II, excluding the stripped SNe~IIb; and interacting SNe, including Types IIn, Ibn, Icn, and Ia-CSM. 
The most well-observed objects of each type in this data release are labeled.
They often are the most nearby or have peculiar evolution. 
Given the frequency of KITS nights and the fact that most of our time allocations are in half nights, we have a maximum of 7 spectra per object, and about 41\% of all objects have only a single spectrum (primarily driven by SNe~Ia).
On average, we obtain 1.9 spectra per object. 
We highlight the SN~Ia~2022exc, SN~Ic-pec~2022jli, SN~II~2022mxv, and SN~Ia-CSM~2022erq, which have 5, 7 (6 included in this release), 6, and 6 spectra, respectively.
These objects are particularly nearby and bright, and show peculiar evolution that warrants multi-epoch observations. 

Figure~\ref{fig:redshift_distribution} displays redshift distributions for each broad class mentioned above. 
Similar to the last figure, the full sample is shown unfilled with DR1 objects in the shaded area. 
The median redshifts are 0.0298 for SNe~Ia, 0.03 for SESNe, 0.0094 for SNe~II, and 0.0355 for interacting SNe (full sample). 
The maximum redshifts are 0.064 for SNe~Ia (SN\,2022cvr), 0.064 for SESNe (SN\,2023mee), 0.03 for SNe~II (SN\,2022dml), and 0.08 for interacting SNe (SN\,2023gpw).
Because of the overabundance of SNe~Ia in the flux-limited sample, we focused on those objects that could either potentially have their luminosity calibrated through direct distance measurements (i.e., Cepheid or tip-of-the-red-giant-branch measurements, corresponding to $D \lesssim 40$~Mpc) or those clearly in the Hubble flow ($z > 0.015$).  
Nevertheless, our choices for the flux- and volume-limited sample result in almost all normal SNe~Ia having $z < 0.04$, which would correspond to the ``Physics'' subsample of CSP~II \citep{hsiao2019}.

We use public photometry of each transient to determine the phase of the NIR spectra.
We primarily use $o$-band ATLAS photometry as the vast majority of our targets are well observed.
For the SNe~II, TDEs, and LRNe that lack a clear time of maximum brightness, we define the phase relative to the ``time of explosion,'' which we define as the midpoint between the first detection and the last nondetection.  
For all other transients, we fit a low-order polynomial to the light curve to determine the peak epoch, from which we define their phase. 
However, there are a few exceptions to these general prescriptions.
SN\,2022jli was discovered soon after it rose for the observing season, indicating that it likely exploded when it was behind the Sun; thus, the phase is relative to the discovery epoch \citep{monard2022}.
SNe\,2022hsu, 2022ihx, and 2022kla have poor ATLAS observations, and we use the ZTF $r$ light curve instead to define phase.
SN\,2022mji set soon after discovery, and thus has a poorly sampled light curve near discovery; we use the last nondetection from ATLAS and the first detection from {\it Gaia} to determine the explosion epoch. 
Table~\ref{tab:KITS_SNe} provides the reference epoch and method for each transient; spectral phases are relative to this reference epoch. 

Figure~\ref{fig:phase_distribution} shows distributions of phases of KITS spectra for each common SN subtype. 
We only display data from the DR1, corresponding to the transients for which we have measured a reference epoch.
The median phases of our observation in DR1 are 12~days post-peak for SNe~Ia, 11 days post-peak for SESNe, 33 days post-explosion for SNe~II, and 26 days post-peak for interacting SNe. 
The minimum phases are 8~days before peak for SNe~Ia (SN\,2022exc); 14~days before peak for SESNe (SN\,2022crv); 1~day post explosion for SNe~II (SN\,2022jzc, ToO observation); and 45~days before peak for interacting SNe (SN\,2022esa).
The maximum phases are 95~days post peak for SNe~Ia (SN\,2022exc), 135 days post-peak for SESNe (SN\,2022jli), 93 days post-explosion for SNe~II (SN\,2022mxv), and 107 days post-peak for interacting SNe (SN\,2022erq).

\subsection{Observations of SNe Ia}

The small extant sample of SNe~Ia from before the start of KITS with high-quality, publicly available NIR spectra is only $\sim 50$~SNe, and with only $\sim 9$ SNe with well-sampled optical photometry appropriate for building a spectral model.  
Most of these spectra are from an unsystematic ensemble of SNe discovered through last-generation targeted SN searches and obtained in 2002--2005 with the 3-meter IRTF \citep{marion2009}. %
The CSP~II \citep{hsiao2019, lu2023} obtained 331 NIR spectra of 94 normal SNe~Ia with light curves and minimal host-galaxy contamination \citep{lu2023}, dramatically increasing the available spectra for training; however, while the spectra have been publicly released, the corresponding light curves have not, preventing the inclusion of CSP~II data in model training samples at this time.

\begin{figure*}
    \centering
    \includegraphics[width=\textwidth,trim={.25cm .25cm .25cm .25cm},clip]{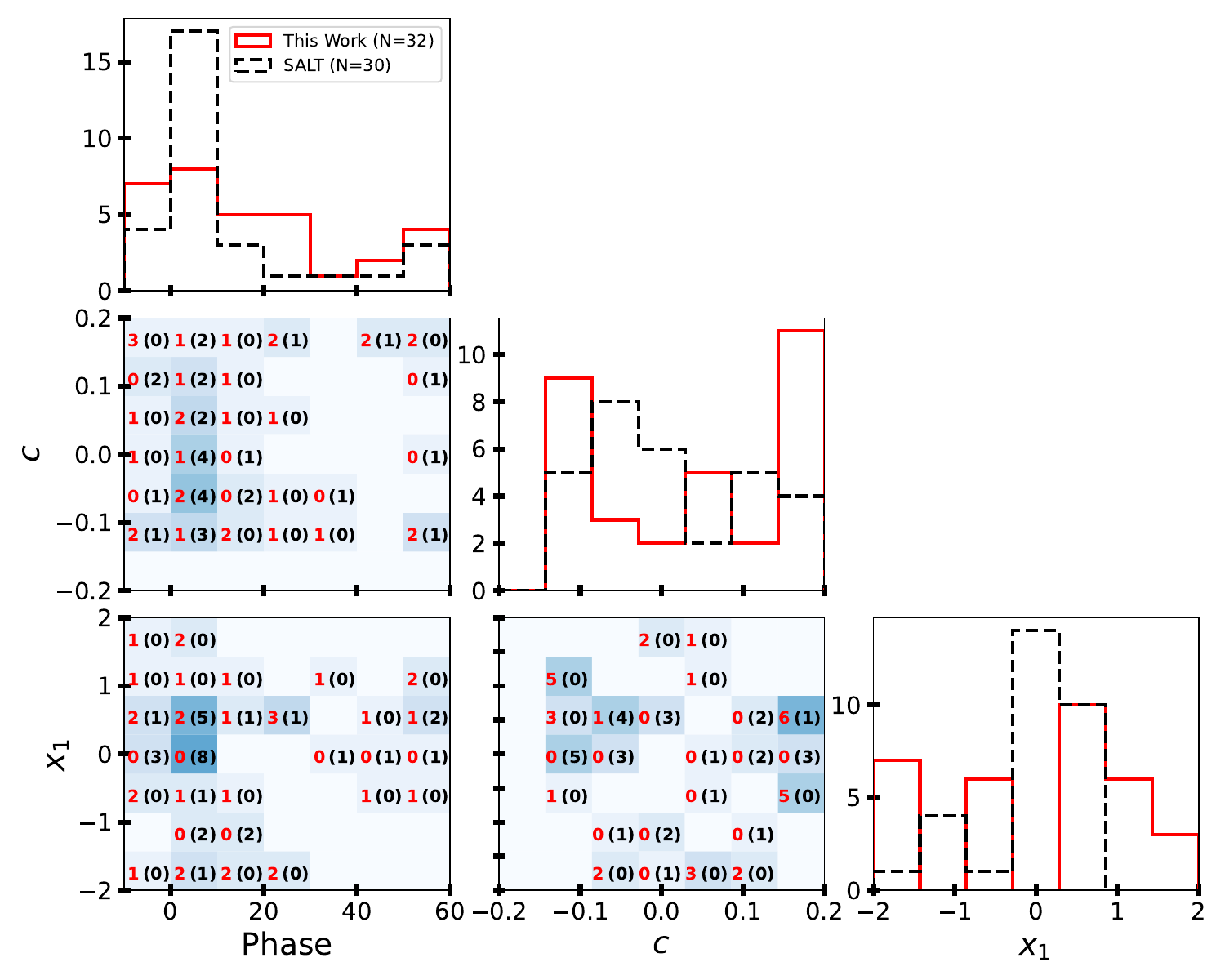}
    \caption{Relationships between SALT3 model parameters for the sample of SN~Ia spectra with coverage beyond $\sim 1.2~\mu$m used in the training of \citet{Pierel22} (black) and those presented in this work (red). 2D histograms show the combined sample in blue, with the individual contributions given as colored numbers.}
    \label{fig:salt_comp}
\end{figure*}

Despite the promise of the NIR, the lack of a robust NIR spectral model means that current cosmological SN constraints exclusively use optical data \citep{Scolnic18:ps1, Riess19, Abbott19:dessn}.  
\citet{Pierel18} estimate that a sample of $\gtrsim 250$ SN~Ia NIR spectra is necessary to include NIR data in dark-energy analyses.  
In addition to a large number of spectra, the SNe must also span the full parameter space of SN properties, requiring coverage in the combination of light-curve shape (e.g., $x_{1}$ for SALT3), color (e.g., $c$ for SALT3), and phase.  
Figure~\ref{fig:salt_comp} displays the KITS coverage in these three parameters compared to the training sample of \citet{Pierel22}, considering only spectra with coverage beyond $\sim 1.2$~$\mu$m.

\section{The data}\label{sec:data}

\begin{figure}
    \centering
    \includegraphics[width=1\linewidth]{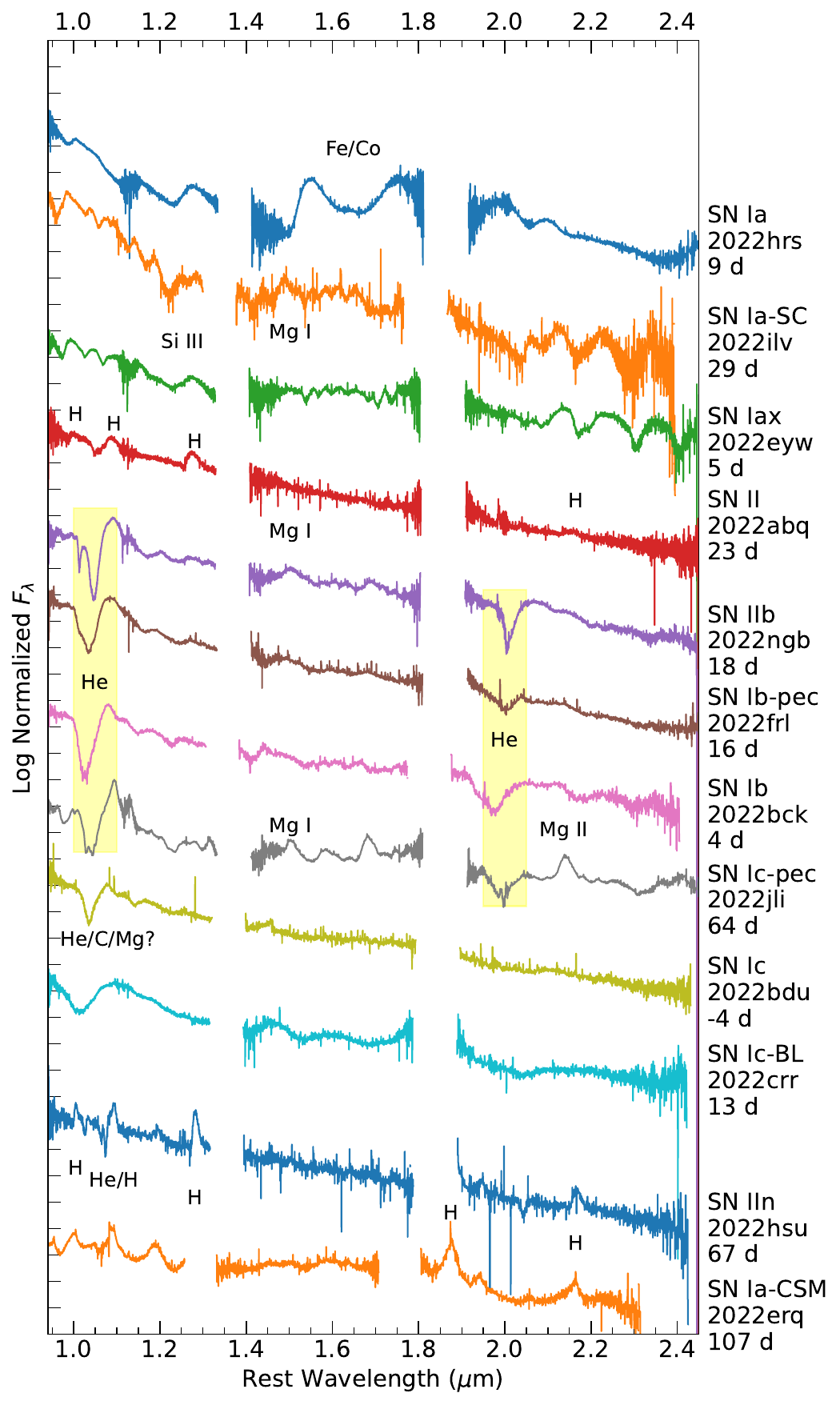}
    \caption{KITS spectra of different subtypes of SNe available in DR1. From top to bottom, we show different subtypes of thermonuclear explosions: normal SN~Ia 2022hrs, a super-Chandra SN 2022ilv, and a Type Iax SN 2022eyq. Then we show different subtypes of CCSNe sorted by the level of stripping: from a Type II SN 2022abq; a partially-stripped Type IIb SN 2022ngb; a peculiar SN Ib 2022frl with some sign of hydrogen; a hydrogen-free Type Ib SN 2022bck; a peculiar Type Ic SN 2022jli with some signature of helium; a helium-poor Type Ic SN 2022bdu; and an energetic Type Ic-BL 2022crr. Finally, we show two interacting SNe: SN IIn 2022hsu and SN Ia-CSM 2022erq. Prominent hydrogen and helium lines are annotated in the plot. Most features in SN~Ia spectra are due to iron and cobalt.}
    \label{fig:all_types}
\end{figure}

Spectra in DR1 are posted to WISEReP for easy access to the time-domain astronomy community.
More data-reduction products are available via Zenodo \citep{tinyanont_zenodo_DR1} to ensure that the final reduction is reproducible.
Extracted and coadded one-dimensional spectra of both the transients and telluric standard stars from \texttt{Pypeit} are provided in FITS format.
This allows users to repeat the flux and telluric calibration if necessary. 

\begin{figure*}
    \centering
    \includegraphics[width=\linewidth]{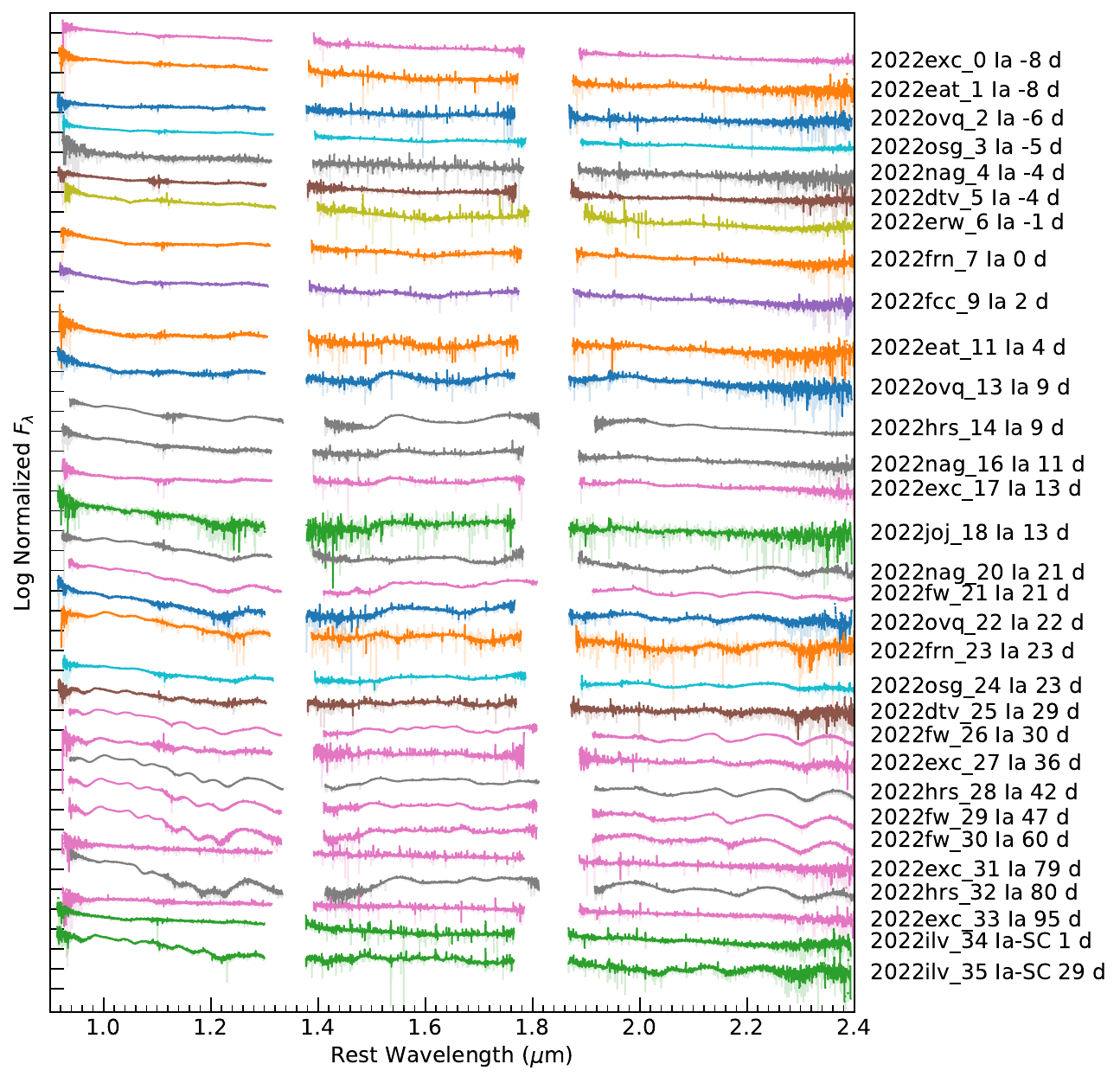}
    \caption{KITS spectra of SNe~Ia included in DR1 sorted by phase of observation. SN name, subtype, and phase are labeled. The spectra are shown in $F_\lambda$ units, normalized and on a log scale. The super-Chandrasekhar event SN\,2022ilv is plotted in the bottom.
    Spectra having poor signal-to-noise ratios are excluded. Some spectra have been smoothed, with the unsmoothed version shown in the background.
    We note that late-time spectra of SN\,2022exc may have significant host-galaxy contamination as the SN is in a nuclear region.
    }
    \label{fig:Ia_seq}
\end{figure*}

\begin{figure*}
    \centering
    \includegraphics[width=\linewidth]{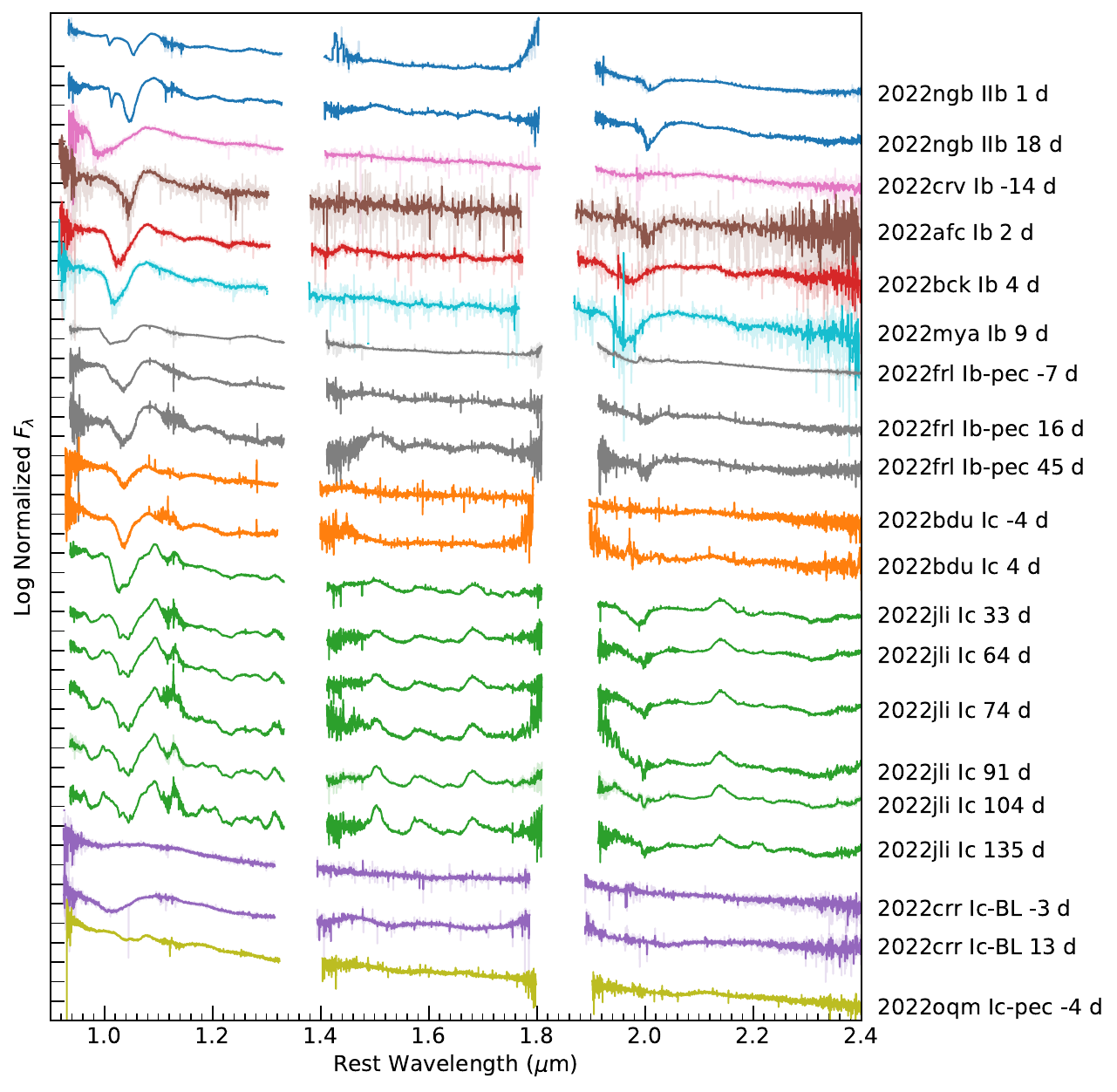}
    \caption{Same as Figure~\ref{fig:Ia_seq} but for SESNe. 
    We order subtypes according to an increasing degree of stripping: hydrogen-poor type IIb, hydrogen-free Ib, and hydrogen-free and helium-poor Ic. 
    The Ic-BL and peculiar Ic are plotted in the bottom.}
    \label{fig:SE_seq}
\end{figure*}

\begin{figure*}
    \centering
    \includegraphics[width=\linewidth]{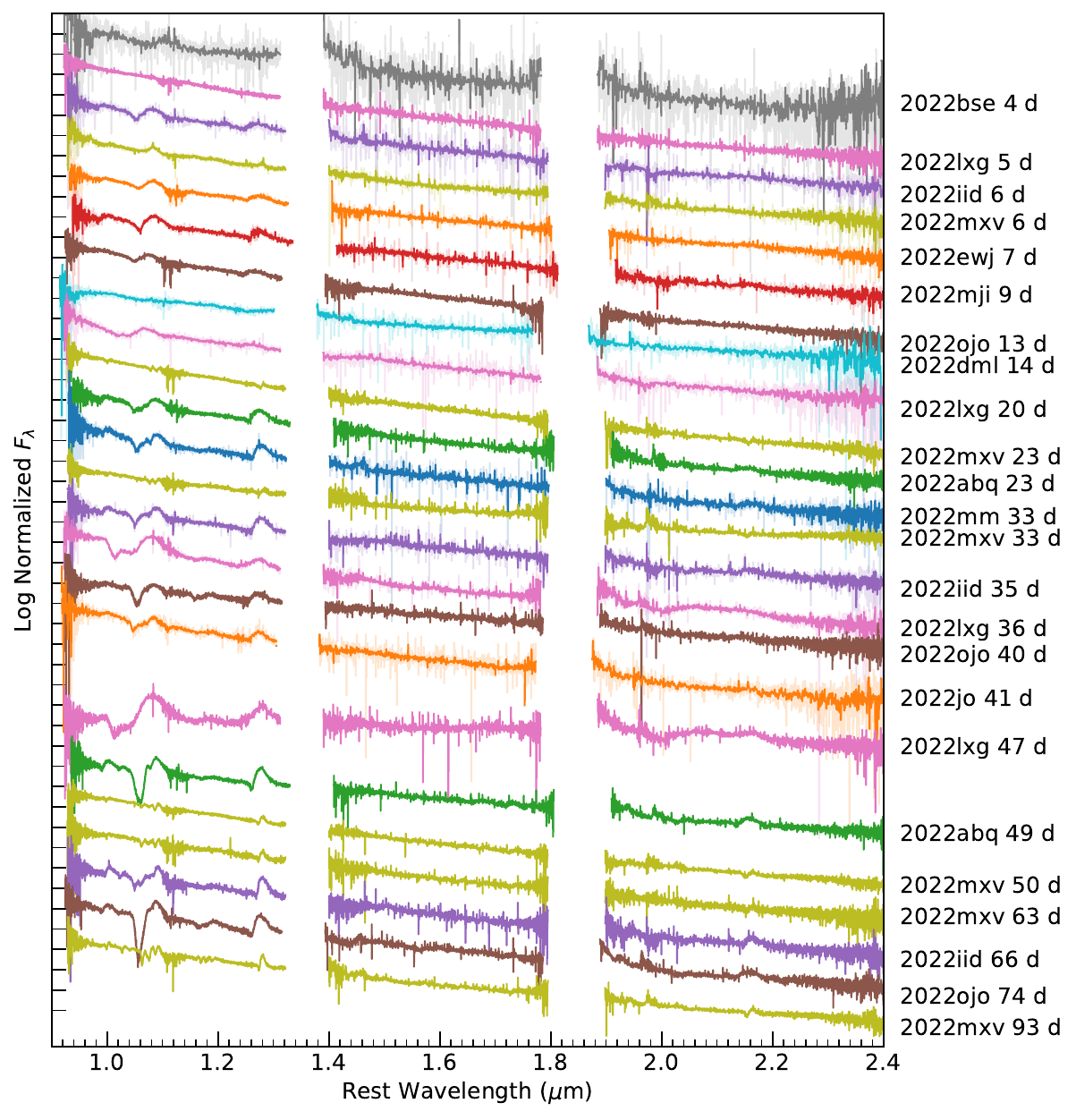}
    \caption{Same as Figure~\ref{fig:Ia_seq} but for SNe~II.}
    \label{fig:II_seq}
\end{figure*}

\begin{figure*}
    \centering
    \includegraphics[width=\linewidth]{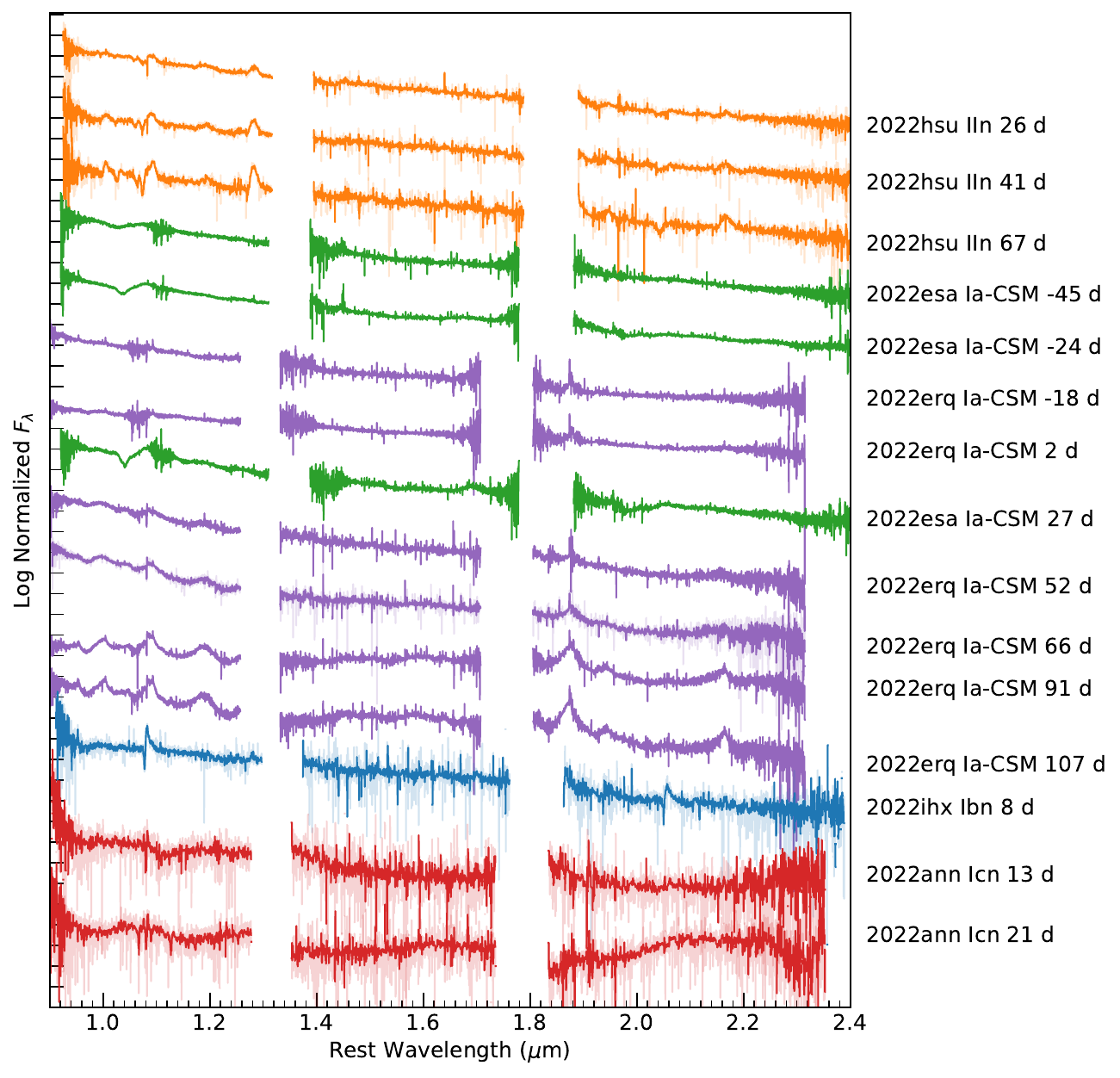}
    \caption{Same as Figure~\ref{fig:Ia_seq} but for interacting SNe.}
    \label{fig:Int_seq}
\end{figure*}

\begin{figure*}
       \centering
    \includegraphics[width=\linewidth]{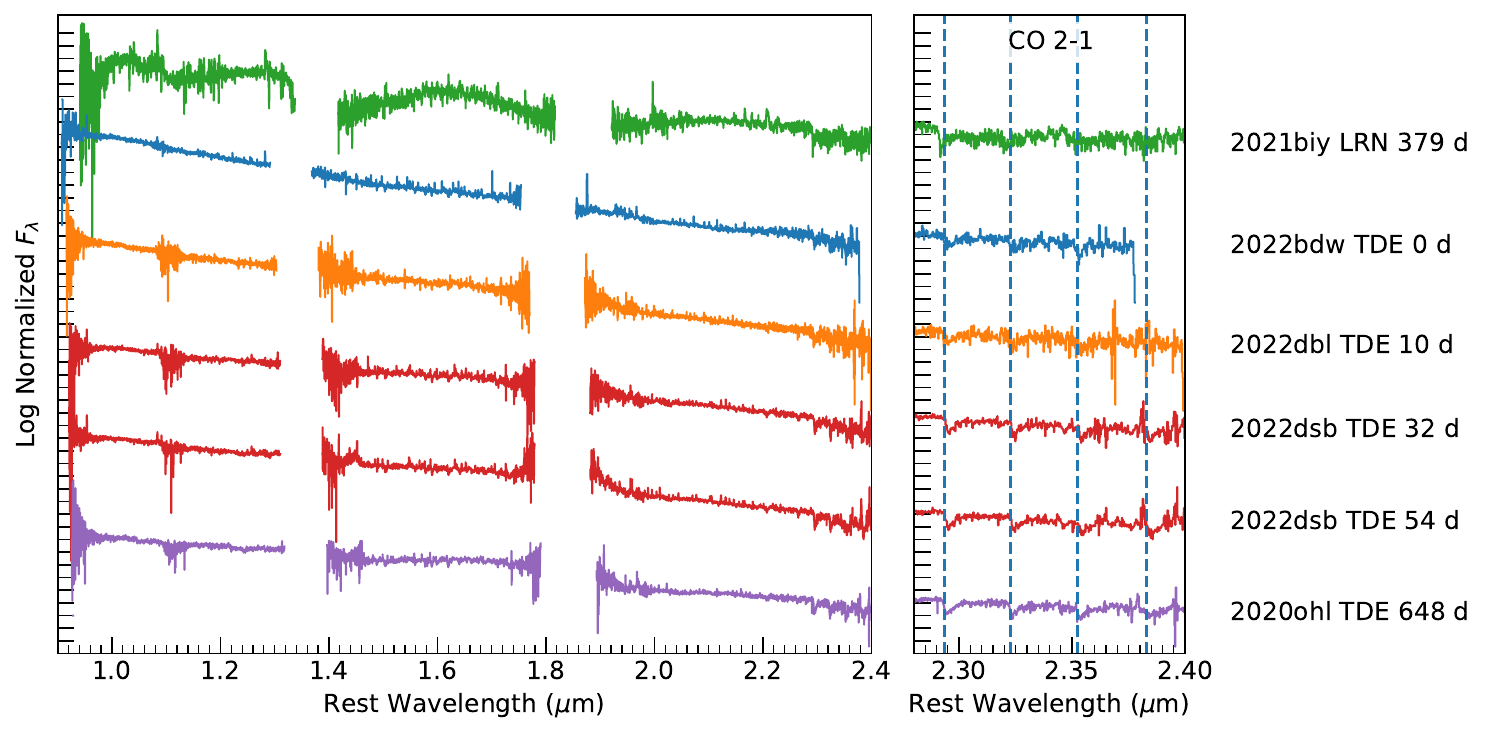}
    \caption{Similar to Figure~\ref{fig:Ia_seq} but for the LRN AT\,2021biy and the TDEs in the sample. The right panel zooms in around the bandheads of the carbon monoxide first overtone to demonstrate the absorption detected in all objects. In the LRN, this could be local to the explosion, while for the TDEs this is likely due to other stars in the galactic nucleus. }
    \label{fig:non_SNe} 
\end{figure*}

To showcase our dataset, we plot representative spectra of different subclasses of SNe available in this data release in Figure~\ref{fig:all_types}, with important spectral lines labeled. 
We also plot most of the observed spectra from DR1 of SNe~Ia, SESNe, SNe~II, and interacting SNe in Figures~\ref{fig:Ia_seq}, \ref{fig:SE_seq}, \ref{fig:II_seq}, and \ref{fig:Int_seq}, respectively.
In addition to SNe, we plot spectra of TDEs and the LRN AT\,2021biy in Figure~\ref{fig:non_SNe}. 
For TDEs, broad hydrogen features seen in the Balmer lines are not visible in the Paschen or Brackett lines for the spectra taken at similar epochs as the optical broad-line detections.

For SNe~Ia, DR1 data only cover normal SNe Ia and one super-Chandrasekhar event (SN\,2022ilv; \citealp{srivastav2023}).
The full KITS data set will cover other subtypes of SNe Ia. 
Spectra shown in Figure~\ref{fig:Ia_seq} show homogeneity, as expected from this class. 
The majority of spectral features seen in the NIR are from iron and cobalt. 
KITS's contribution to SNe~Ia is in the coverage of the light curve parameter space discussed in the previous section. 

For SESNe, this dataset can probe the diminishing NIR helium features from Type IIb to Ib to Ic. 
The 2.0581~$\mu$m line is uncontaminated and unsaturated; thus, it could be used to measure the helium mass in the envelope at the time of core collapse \citep{dessart2020}. 
Our dataset contains three peculiar objects. 
First, SN\,2022frl is classified as a Type Ib SN; 
however, its NIR spectra contain clear hydrogen emission at all phases, resembling those of Type IIb SN\,2022ngb (Figure~\ref{fig:SE_seq}). 
This indicated that SN\,2022frl may have an ambiguous classification between Types IIb and Ib.  
SN\,2022jli is classified as Type Ic in the optical, but NIR spectra show clear absorption associated with the \ion{He}{1} 2.0581~$\mu$m line. 
The absorption trough of the \ion{He}{1} 1.083~$\mu$m line also has a multicomponent structure with several distinct absorption troughs, which may indicate inhomogeneous ejecta.
The last peculiar object is SN\,2022oqm, which is more similar to calcium-rich transients than SNe \citep{karthik2023}.

For SNe~II, most data obtained with KITS (and all that are included in DR1) are during the plateau phase. 
The highlight of this dataset is the diverse absorption profile of the \ion{He}{1} 1.0830~$\mu$m and hydrogen Pa$\gamma$ (1.0938~$\mu$m) complex. 
Some objects, such as SNe\,2022lxg and 2022iid, show strong multicomponent absorption with a high-velocity component associated with CSM interaction \citep{chugai2007}. 
We note that SN IIb 2022ngb also has a very strong high-velocity absorption, likely of both \ion{He}{1} 1.0830~$\mu$m and Pa$\gamma$.
Others, such as SNe\,2022abq and 2022ojo, have a more typical P~Cygni profile. 
We also noticed a peculiar spectral evolution for SN\,2022mxv (which was classified using a KITS spectrum; \citealp{davis2022mxv}), a luminous SN~II with clear narrow emission lines from CSM interaction at early times.
This is the only object for which the velocity remains low throughout its evolution such that the \ion{He}{1} 1.0830~$\mu$m and Pa$\gamma$ are clearly separated for all phases. 
This indicates that the photosphere remains above the higher-velocity ejecta, and only the shocked CSM is observed. 
Another peculiar object is SN\,2022lxg, which has a distinct profile for hydrogen lines, and flux excess in the $K$ band starting around 37~days post-explosion. 
This is normally far too early for an SN~II to have dust form in the ejecta \citep[e.g.,][]{sarangi2018}, but with the sign of CSM interaction detected at early times, we might be seeing CSM dust being heated. 
The fact that so many peculiar objects are identified in the limited set of objects included in DR1 highlights the relatively unexplored nature of NIR spectroscopy of transients. 

For interacting SNe, the NIR allows for observations of isolated helium lines. 
As we discuss below, this leads to a discovery that the helium-poor interacting Type Icn SNe have a small amount of helium in the CSM. 
This data release also contains multi-epoch observations of two SNe~Ia-CSM: SNe\,2022erq and 2022esa.
They are thought to be SNe~Ia exploding in a dense hydrogen-rich medium. 
The NIR spectra of SN\,2022erq, in particular, show complicated line profiles for the \ion{He}{1}~1.0830~$\mu$m and Pa$\gamma$ complex, with very narrow absorption and emission components from both lines superimposed on broad electron-scattered lines.

KITS data have already led to five publications at the time of writing. 
\citet{davis2023} presented KITS NIR spectroscopy of SN\,2022ann, a member of the newly established class of hydrogen-free and helium-poor interacting SNe~Icn, along with other optical observations. They found that the \ion{He}{1} 1.083~$\mu$m line is unambiguously present in KITS NIR spectra at phases where the corresponding optical lines are not identifiable.
Owing to the small velocities in the CSM, this helium line is not blended with the \ion{C}{1} 1.0693~$\mu$m line, which is also detected. With these observations, they conclude that NIR spectra provide stronger constraints on the presence of helium in SNe~Icn than optical-only datasets.
\citet{karthik2023} presented KITS NIR spectroscopy of SN\,2022oqm, a peculiarly luminous calcium-rich transient (CaRTs). 
The nondetection of the \ion{He}{1} 2.0581~$\mu$m line in the KITS NIR spectrum was used to argue that SN\,2022oqm has a particularly helium-poor atmosphere, unlike other CaRTs, which normally have strong helium features. This allowed for important constraints on the progenitor system. The obtained KITS NIR spectrum is also one of the earliest NIR spectra of a CaRT, allowing for a better probe of the early-time NIR evolution.
\citet{padilla2023} used KITS data to explore the presence of unburned helium by detecting the \ion{He}{1}~1.0830~$\mu$m line in SN\,2022joj. 
Recent simulations have highlighted the potential occurrence of unburned helium in both single and double detonations within the outer ejecta. 
Consequently, given the indications that SN\,2022joj exhibits characteristics consistent with a potential double detonation event, the investigation of this line becomes pivotal.
Model comparison to the KITS spectrum constrained the helium shell mass to around 0.02~$M_\odot$.
\citet{siebert2023} presented KITS nebular spectroscopy of SN\,2022pul (a ``super-Chandrasekhar"-mass SN~Ia) along with nearly simultaneous NIR and MIR spectroscopy from {\it JWST}. 
The higher resolution of the KITS data in the NIR was critical for constraining velocity distributions of IGEs in the SN ejecta. 
\citet{kwok2023} found that IGE emission-line profiles tended to be redshifted while intermediate-mass element emission-line profiles were blueshifted. 
This supported their conclusion that SN\,2022pul was the result of the violent merger of two white dwarfs. 
Additionally, given that the {\it JWST} data reveal a strong thermal dust continuum in SN\,2022pul \citep{siebert2023}, Johansson et al. (2023, in prep.) further analyzed the KITS data to constrain the presence of CO emission.

\section{Summary}\label{sec:summary}
The first KITS data release included in this work consists of 105 NIR spectra from 50 transients. 
We aim to provide KITS data in a timely manner, with the data released 1--2~yr after they were obtained.  To meet this fast timeline, the data release is limited in scope.  Future data releases will include additional NIRES spectra, NIRES NIR photometry, NIR spectra from IRTF/SpeX and SOAR/TripleSpec, Gemini/FLAMINGOS2 photometry, and photometry and spectroscopy from other sources.

The KITS survey design combined with unbiased discovery surveys resulted in a diverse set of astrophysical transients beyond previous NIR datasets that existed in the literature.
The last large-scale NIR spectroscopic survey of SNe, CSP-II, concluded in 2015, and KITS provides NIR spectra of a contemporary sample of transients that include newly discovered classes of objects.
This dataset will serve as a stepping stone to analyze \textit{JWST} observations of high-redshift transients, and to plan the time-domain survey for \textit{Roman} for next-generation cosmological studies using Type Ia SNe in the NIR. 
With the full data release expected in 2024, KITS data will account for a significant portion of NIR spectra of transients.
This will contribute to the NIR spectroscopic treasury, against which new observations can be compared to reveal patterns and features previously invisible to us.

\section*{Acknowledgements}
NASA Keck time is administered by the NASA Exoplanet Science Institute. Data presented herein were obtained at the W.\ M.\ Keck Observatory from telescope time allocated to the National Aeronautics and Space Administration through the agency's scientific partnership with the California Institute of Technology and the University of California. The Observatory was made possible by the generous financial support of the W.\ M.\ Keck Foundation.

The authors wish to recognize and acknowledge the very significant cultural role and reverence that the summit of Mauna Kea has always had within the indigenous Hawaiian community. We are most fortunate to have the opportunity to conduct observations from this mountain.

We thank Keck Observatory support astronomers and staff, especially P.\ Gomez and J.\ Walawender, for assisting us acquire data published in this work.
We are grateful to J.\ X.\ Prochaska, J.\ Hennawi, and F.\ Davies for helping us understand \texttt{Pypeit} and troubleshooting reduction issues.

The Keck Infrared Transient Survey was executed primarily by members of the UC Santa Cruz transients team, who were supported in part by NASA grants NNG17PX03C, 80NSSC21K2076, 80NSSC22K1513, 80NSSC22K1518; NSF grant AST--1911206; and by fellowships from the Alfred P.\ Sloan Foundation and the David and Lucile Packard Foundation to R.J.F.  KITS was directly supported by NASA grant 80NSSC23K0301.

C.D.K.\ is partly supported by a CIERA postdoctoral fellowship.
S.T.\ was supported by the Cambridge Centre for Doctoral Training in Data-Intensive Science funded by STFC.
Support for J.R.P. was provided through NASA Hubble Fellowship grant HF2-51541.001-A, awarded by the Space Telescope Science Institute (STScI), which is operated by
the Association of Universities for Research in Astronomy, Inc., under NASA contract NAS5-26555.
L.G.\ acknowledges financial support from the Spanish Ministerio de Ciencia e Innovaci\'on (MCIN), the Agencia Estatal de Investigaci\'on (AEI) 10.13039/501100011033, and the European Social Fund (ESF) ``Investing in your future" under the 2019 Ram\'on y Cajal program RYC2019-027683-I and the PID2020-115253GA-I00 HOSTFLOWS project, from Centro Superior de Investigaciones Cient\'ificas (CSIC) under the PIE project 20215AT016, and the program Unidad de Excelencia Mar\'ia de Maeztu CEX2020-001058-M.
C.L. acknowledges support from the National Science Foundation (NSF) Graduate Research Fellowship under grant DGE-2233066.
M.R.S.\ is supported by an STScI Postdoctoral Fellowship.
A.V.F.\ is grateful for financial support from the Christopher R. Redlich Fund and many other donors.
W.J-G is supported by the National Science Foundation Graduate Research Fellowship Program under Grant No.~DGE-1842165. W.J-G acknowledges support through NASA grants in support of {\it Hubble Space Telescope} program GO-16075 and 16500.
J.S.B. acknowledges support from the Gordon and Betty Moore Foundation.
S.M.W was supported by the UK Science and Technology Facilities Council (STFC).
This work was supported in part by the Director, Office of Science, Office of High Energy Physics of the U.S. Department of Energy under Contract No. DE-AC02-05CH11231.
K.S.M. and M.G. acknowledge funding from the European Union’s Horizon 2020 research and innovation programme under ERC Grant Agreement No. 101002652.
A.V.F.'s team at UC Berkeley is grateful for financial assistance from the Christopher R. Redlich Fund and many individual donors.

This research has made use of the Keck Observatory Archive (KOA), which is operated by the W. M. Keck Observatory and the NASA Exoplanet Science Institute (NExScI), under contract with NASA.
Part of this work uses the Chalawan High-Performance Computer Cluster at the National Astronomical Research Institute of Thailand (NARIT).

\bibliography{KITS}{}
\bibliographystyle{aasjournal}

\end{document}